\documentclass{article}

\usepackage{arxiv}
\usepackage{float}
\usepackage{amsmath}
\usepackage[utf8]{inputenc} % allow utf-8 input
\usepackage[T1]{fontenc}    % use 8-bit T1 fonts
\usepackage{hyperref}       % hyperlinks
\usepackage{url}            % simple URL typesetting
\usepackage{booktabs}       % professional-quality tables
\usepackage{amsfonts}       % blackboard math symbols
\usepackage{nicefrac}       % compact symbols for 1/2, etc.
\usepackage{microtype}      % microtypography
\usepackage{lipsum}		% Can be removed after putting your text content
\usepackage{graphicx}
\usepackage[compress]{natbib}
\usepackage{doi}

\title{Temperature Dependent Mechanical and Structural Properties of Uniaxially Strained Planar Graphene}
\date{October 2025}	% Here you can change the date presented in the paper title
\author{ \href{https://orcid.org/0000-0001-6582-9117}{\includegraphics[scale=0.06]{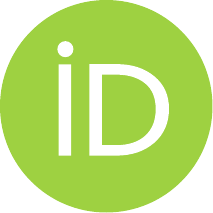}\hspace{1mm}San\'e Erasmus}\\
	Nonlinear Dynamics and Chaos Group\\
	Department of Mathematics and Applied Mathematics\\
	University of Cape Town\\
	Rondebosch 7700, South Africa\\
	\And
	\href{https://orcid.org/0000-0002-2276-8251}{\includegraphics[scale=0.06]{orcid.pdf}\hspace{1mm}Charalampos Skokos} \\
	Nonlinear Dynamics and Chaos Group\\
	Department of Mathematics and Applied Mathematics\\
	University of Cape Town\\
	Rondebosch 7700, South Africa\\
	\&\\
	Max Planck Institute for the Physics of Complex Systems\\
	N\"othnitzer Str.~38, D-01187 Dresden, Germany\\
	\And
	\href{https://orcid.org/0000-0001-7763-718X}{\includegraphics[scale=0.06]{orcid.pdf}\hspace{1mm}George Kalosakas}\thanks{Correspondence: georgek@upatras.gr} \\
	Department of Materials Science\\
	University of Patras\\
	GR-26504 Rio, Greece\\
}

\renewcommand{\headeright}{}
\renewcommand{\undertitle}{Preprint}
\hypersetup{
	pdftitle={Temperature Dependent Mechanical and Structural Properties of Uniaxially Strained Planar Graphene},
	pdfauthor={San\'e Erasmus, Charalampos Skokos, George Kalosakas},
	pdfkeywords={graphene, molecular dynamics, stress-strain response, elastic properties, bond length and angle distributions},
}

\begin{document}
	\maketitle
	
	\begin{abstract}
		Using molecular dynamics simulations in a planar graphene sheet, we investigate the temperature dependence of its mechanical behavior under uniaxial tensile stress applied either along the armchair or the zigzag direction. Stress-strain curves are calculated for different temperatures and the corresponding dependence of various elastic parameters, 
		%like the Young modulus, the third-order elastic modulus, the tensile strength and failure strain,
		is discussed. Fracture stress and strain, as well as the Young modulus, decrease almost linearly with temperature, {in accordance with previous investigations. An almost linear variation of the third-order elastic modulus with temperature is demonstrated, revealing opposite trends for uniaxial loadings along the armchair or the zigzag direction. The detailed dependence of the distributions of bond lengths and bond angles both on strain and temperature is presented for the first time, along with approximate analytical expressions.} The latter describe accurately the numerically obtained distributions.  
	\end{abstract}

	% keywords can be removed

	%%%%%%%%%%%%%%%%%%%%%%%%%%%%%%%%%%%%%%%%%%
\section{Introduction}

Since the discovery of graphene, there have been a number of investigations into its mechanical behavior. Despite the difficulty of applying controlled mechanical loads at the nanoscale, experimental studies have verified an exceptional value of stiffness and extremely high tensile strength \cite{ref-Lee-Science-2008,ref-Cao-NatComm-2020, ref-Varillas-MechSci-2024}, in accordance with corresponding theoretical predictions. {For example,  in Reference \cite{ref-Lee-Science-2008} the mechanical properties of graphene were deduced from nanoindentation experiments on circularly clamped samples, while in References \cite{ref-Cao-NatComm-2020,ref-Varillas-MechSci-2024} push-to-pull testing was employed to directly apply uniaxial strain.} There exist a number of related numerical investigations using molecular dynamics (MD) simulations with a variety of potential functions \cite{ref-Zakharchenko-PRL-2009,ref-Zhao-NanoLet-2009,ref-Neek-Amal-PRB-2010,ref-Shen-MatDes-2010,ref-Tsai-MatDes-2010,ref-Zhao-JAP-2010,ref-Wei-JCP-2011,ref-Dmitriev-CompMatSci-2012,ref-Kalosakas-JAP-2013,ref-Baimova-EUPhyLet-2013,ref-Zhang-CompMatSci-2013,ref-Sgouros-2DMat-2016,ref-Sgouros-SciRep-2018,ref-Lebedeva-PhysE-2019,ref-Savin-MechofMat-2019,ref-Aparicio-Nanotech-2021,ref-deSousa-Intech-2021,ref-Kalosakas-Mat-2021,ref-Flores-PhysChem-2022,ref-Fthenakis-FrontChem-2022,ref-Varma-IntJMechMatDes-2022,ref-Pacheco-Sanjuan-Carbon-2023,ref-Sgouros-PRE-2024}, density functional theory \cite{ref-Kudin-PRB-2001,ref-Liu-PRB-2007,ref-Zeinalipour-Yazdi-JAP-2009,ref-Marianetti-PRL-2010,ref-Wagner-PRB-2011,ref-Mirnezhad-JTS-2012,ref-Shao-JCP-2012}, or other theoretical approaches including molecular mechanics \cite{ref-Lu-IJAM-2009,ref-Fthenakis-JoP-2017,ref-Genoese-FrontMat-2019}, and combinations of continuum elasticity theory with other methods \cite{ref-Cadelano-PRL-2009,ref-Samadikhah-CompMatSci-2012,ref-Wei-NanoLet-2013,ref-Tanhadoust-DiamRelMat-2022}. 

The temperature dependence of various elastic properties of graphene has been examined using MD~\cite{ref-Shen-MatDes-2010,ref-Zhao-JAP-2010,ref-deSousa-Intech-2021}, Monte Carlo atomistic simulations~\cite{ref-Zakharchenko-PRL-2009}, density functional theory \cite{ref-Shao-JCP-2012}, or the asymptotic homogenization method \cite{ref-Tahani-AMM-2024}. 
Moreover, a few MD studies have investigated the variation of bond lengths and bond angles with uniaxial tensile loading \cite{ref-Zhao-NanoLet-2009, ref-Kalosakas-Mat-2022}. The latter work shows results from first principles methods too, and presents analytical expressions for the dependence of these structural parameters on strain~\cite{ref-Kalosakas-Mat-2022}. The variation of bond lengths and angles with biaxial strain has been discussed in \cite{ref-Dheeraj-Nanotech-2020}. {To our knowledge, the dependence of the  distributions of bond lengths and bond angles on both temperature and strain has not yet been examined. }

In the present work, we use MD simulations to study the behavior of planar graphene under uniaxial tensile load, considering the influence of temperature. In particular, we implement symplectic integration methods for simulating the system's time evolution, which allow highly accurate computations for arbitrarily long times. We calculate stress-strain curves at various temperatures and from these results we estimate the variation of several elastic parameters with temperature.
{Our findings for the Young modulus, fracture strength, and failure strain, are in agreement with previous studies. The temperature dependence of the the third-order elastic modulus has not been presented up to now.}

Furthermore, we compute bond lengths and bond angles of bulk graphene over a large time-window after thermal equilibrium has been reached, and subsequently analyze these results in order to get the dependence of the corresponding distributions on both stress and temperature. Finally, we present analytical expressions which closely match the numerically obtained distributions of bond lengths and angles. Thus, we describe, {for the first time to the best of our knowledge,} the detailed dependence of these structural properties of graphene on both the applied tensile stress and temperature. 

The paper is organized as follows. In Section~\ref{section-model} we present the used force field, along with the numerical methods we implement. The results of our investigation are discussed in Section~\ref{section-results}. In particular, the implementation of finite temperatures in our MD microcanonical simulations is discussed in Section~\ref{section-E-T}, and thermal effects on graphene's mechanical response under uniaxial tension are studied in Section~\ref{section-mechanical-response}. Then we present the distribution of bond lengths and bond angles in the planar sheet of $sp^2$ carbon atoms at various stresses and temperatures in Section~\ref{section-distributions}, while analytical expressions for said distributions are determined in Section~\ref{section-fittings}. Finally we conclude our findings in Section~\ref{section-conclusion}.

%%%%%%%%%%%%%%%%%%%%%%%%%%%%%%%%%%%%%%%%%%
\section{Model and numerical methods}
\label{section-model}

\begin{figure}[H]
	\centering
	\includegraphics[width=6cm]{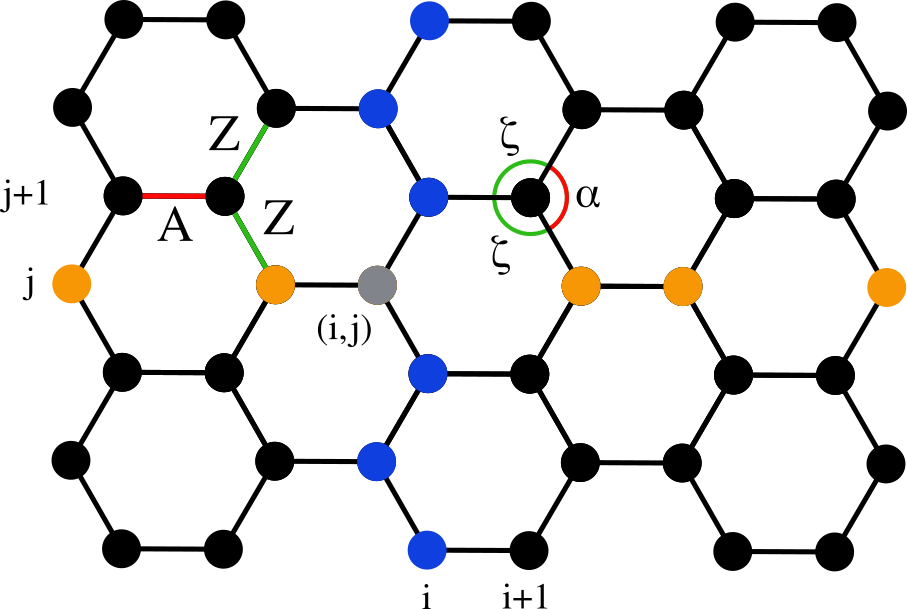}
	\caption{A schematic of the hexagonal graphene lattice depicting $N=42$ atoms, arranged in $N_I=6$ columns and $N_J=7$ rows. Atoms in column $i$ and row $j$ are indicated in blue and orange, respectively, and the $(i,j)$th atom is colored in gray. The $A$ and $Z$ type bonds, and similarly the $\alpha$ and $\zeta$ type angles, are respectively indicated in red and green (see Section~\ref{section-distributions} for more details on these distinctions).}
	\label{fig-lattice}
\end{figure}

We consider a two-dimensional (2D) model of graphene as a hexagonal lattice of carbon atoms within a plane. Figure~\ref{fig-lattice} illustrates a part of this structure at equilibrium, where the distance between any two neighboring atoms is $r_0=1.42$~\AA~and the angle formed by three consecutive atoms is $\phi_0=2\pi/3$~rad. Furthermore, all carbon atoms have mass $m=12$~amu. In the orientation depicted in Figure~\ref{fig-lattice}, the top and bottom edges represent the "{\it armchair edges}", while the left and right edges correspond to the "{\it zigzag edges}". Furthermore, it is common to call "{\it armchair direction}" the horizontal direction in Figure~\ref{fig-lattice}, and "{\it zigzag direction}" the vertical one. 
%============================
%============================

We refer to columns and rows within the graphene sheet, as indicated by the blue and orange colored atoms, respectively, in Figure~\ref{fig-lattice}. A lattice of size $N=N_I\times N_J$ consists of $N_I$ columns, which are indexed by $i$, and $N_J$ rows, which are indexed by $j$. The $(i,j)$th atom is indicated in gray in Figure~\ref{fig-lattice}. Thus, the configuration shown in this figure corresponds to a $N_I\times N_J = 6\times7$ lattice, totaling $N=42$ atoms.
In our simulations we consider a lattice of $N_I=86$ columns and $N_J=87$ rows of atoms, resulting in a $N=86\times87=7482$ total number of carbon atoms. This lattice size is sufficiently large to negate small-size effects~\cite{ref-Kalosakas-JAP-2013}, allowing thus to represent the behavior of bulk graphene, while still permitting extensive numerical simulations within reasonable CPU times. We have further confirmed that indicative data for larger lattices are similar to the ones presented here.
%In our simulations we use lattices where $N_J = N_I+1$,  but the considered numerical model works for any even integer $N_I$ and odd integer $N_J$ in the case of free boundary conditions. For periodic boundary conditions we require both $N_I$ and $N_J$ values to be even. 

A Hamiltonian formalism is used to investigate the in-plane dynamics of the lattice in a similar fashion as that of Reference~\cite{ref-Hillebrand-Chaos-2020}. The atomistic force field, describing bond stretching and angle bending deformations, has been determined through fittings with relevant density functional energy computations \cite{ref-Kalosakas-JAP-2013}. In particular, the potential energy of a covalent bond between nearest neighboring carbon atoms at distance $r$ is given by the Morse expression,
\begin{equation}
	V_M(r) = D\left(e^{-a\left(r-r_0\right)}-1\right)^2,
	\label{eqn-morse}
\end{equation} 
where $D=5.7$ eV is the depth of the potential well and $a = 1.96$ \AA$^{-1}$ is the inverse characteristic length scale of the potential. The angle bending energy term describing a bond angle $\phi$ formed by three consecutive atoms is
\begin{equation}
	V_B(\phi) = \frac{d}{2}\left(\phi-\phi_0\right)^2-\frac{d'}{3}\left(\phi-\phi_0\right)^3,
	\label{eqn-bending}
\end{equation} 
where $d=7.0$~eV/rad$^2$ and $d'=4$~eV/rad$^3$ are the constants of the quadratic and nonlinear term of the potential, respectively. 

{A limitation of the used force field, given by Equations~\eqref{eqn-morse} and \eqref{eqn-bending}, is the decoupling of the bond stretching and angle bending variables. Though this is a standard approximation in atomistic simulations, in real systems these degrees of freedom are coupled.
	However, the predictions of this model for the Young modulus and the intrinsic strength of graphene are in good agreement with the experimental estimates of Reference~\cite{ref-Lee-Science-2008}, while the stress-strain response is in accordance with corresponding calculations from first principles~\cite{ref-Kalosakas-JAP-2013}.} 
{A more drastic approximation is that the out-of-plane atomic displacements are ignored. At finite temperatures ripples and other out-of-plane deformations spontaneously appear in graphene due to thermal fluctuations \cite{ref-Fasolino-NMat-2007,ref-Meyer-Nature-2007}. However, we expect that at relatively low temperatures these non-planar deformations would be suppressed when graphene is uniaxially stressed.}
{Stretching has been experimentally used to flatten graphene \cite{ref-Cao-NatComm-2020} and this has been further confirmed by MD simulations \cite{ref-Lee-NanoResLets-2015}. } 
{Thus, we have considered here only temperatures up to 700~K and not higher ones, even though graphene exists at much larger temperatures.}
{As will be discussed later, the intrinsic strength and fracture strain obtained via our model at room temperature are in good agreement with those obtained from fully three-dimensional MD simulations employing different interatomic potentials \cite{ref-Zhao-NanoLet-2009,ref-Lee-NanoResLets-2015,ref-Varillas-MechSci-2024}.}

The total energy of the system (i.e., the values of the model's Hamiltonian $H$) is the sum of the above potential energy terms for all bond lengths between nearest neighboring atoms and all bond angles between adjacent bonds, and the kinetic energy of each atom. Denoting the total potential energy at time $t$ by $E_V(t)$ and the total kinetic energy at $t$ by $E_K(t)$, the Hamiltonian 
\begin{equation}
	H = E_K(t) + E_V(t),
	\label{eq:total_energy}
\end{equation}
%Denoting the energy of atom $(i,j)$ at time $t$ by  $h(i,j)(t)$ the Hamiltonian \eqref{eq:total_energy} is written as
% \begin{equation}
	%     H(t)=\sum_{j=0}^{N_J-1}\sum_{i=0}^{N_I-1} h(i,j)(t).
	%     \label{eq:hamiltonian_sum}
	% \end{equation}
%In this expression $h(i,j)(t)$ is obtained from half the bond length potential of each bond of atom $(i,j)$, the bond angle potential for each angle formed about atom $(i,j)$ by adjacent atoms, and the kinetic energy of atom $(i,j)$ (for more details see \cite{ref-Ngapasare-MSc-2017,ref-Hillebrand-Chaos-2020}). We determine the expressions for each $h(i,j)(t)$ in terms of
is expressed through the positions $(x(t),y(t))$ and the corresponding conjugate momenta of all carbon atoms within the considered graphene sheet. The time evolution of each atom's position and momentum is governed by the system's Hamilton's equations of motion, which conserve the total energy, Equation~\eqref{eq:total_energy}. 

To investigate the effects of uniaxial tensile load, a constant force $f$ is applied to all atoms on the appropriate edges of the sheet \cite{ref-Kalosakas-JAP-2013,ref-Kalosakas-Mat-2022}: For stress/strain along the armchair direction, the force $f$ is applied on the atoms of the zigzag edges, where on the opposite edges opposite forces, directed outwards, are applied. Similarly, for stress/strain along the zigzag direction, the force $f$ is applied to the armchair edges, again with opposite forces on opposite edges. Tensile loading results in additional terms in the Hamiltonian, given by appropriate products of the relevant edge atom displacements with the applied force $f$. For constant forces, as in our case here, the conservation of the system's total energy still holds.

{We emphasize that we perform stress-controlled simulations here, where we fix the forces (stress) and directly compute the resulting strain. This is a natural choice in MD, in contrast to imposing fixed displacements (strain-controlled simulations) which is preferred in first principles' studies.
	In principle, these two methods of strain- or stress-controlled simulations are equivalent for estimating the stress-strain mechanical response of the system. For example, one can see in Figures 4 and 5 of Reference~\cite{ref-Kalosakas-JAP-2013} the direct comparison of stress-controlled MD data and strain-controlled density functional theory data, which produce identical results at least in the linear response regime.}

In two-dimensional materials like graphene, the stress is expressed as force per unit length. Taking into account the distance between successive atoms at the relevant edges where the force is applied, i.e. the nearest neighboring atoms along an edge column{, or row,} in Figure~\ref{fig-lattice} concerning stress in the armchair{, or zigzag respectively,} direction, the following relations connect nominal stresses and forces 
\begin{equation}
	\sigma_a = \frac{f}{r_0 \sin(\phi_0/2)} \,\,\,\, \mbox{and} \,\,\,\, \sigma_z = \frac{f}{0.5\,r_0\left[1+\cos(\phi_0/2)\right]},
	\label{eq:sigmas}
\end{equation}
where the indices $a$ and $z$ denote stress in the armchair and zigzag direction, respectively, while $r_0$ and $\phi_0$ are the equilibrium values mentioned above.

To determine at zero temperature the relaxed state of the strained graphene for various applied stresses in any direction, a friction term proportional to the velocity of each atom is incorporated in the MD simulations, setting the friction coefficient to $\gamma=0.1$~ps$^{-1}${; see Reference~\cite{ref-Kalosakas-JAP-2013} and the discussion therein.} Then, the fourth order Runge-Kutta numerical integration technique is used with an integration time step of $dt=0.005~t_u$, where $t_u=0.0102$~ps represents the time unit in our model. This time step ensures that the relative energy error $Err(t)=\left|H(t)-H(0)\right|/\left|H(0)\right|$ is below $10^{-7}$ in corresponding energy conserving simulations {where the friction term is absent.}
However, we now simulate the dynamics of the dissipative version of the system until times $t_f=3\times10^3~t_u$, when the total kinetic energy is practically zero ($E_K(t_f)<10^{-16}$~eV). In this way we determine the relaxed equilibrium positions of the atoms within the lattice, for each considered value of stress $\sigma$.

Based on this equilibrium configuration of graphene subjected to  tensile loads, we embark on the main phase of the numerical investigation: following the dynamics of the lattice for a fixed value of stress $\sigma$, at various temperatures $T$, for a long enough time to allow deductions about the thermal equilibrium properties of the stressed material. For these numerical simulations we implement the symplectic integrator ABA864 \cite{ref-Blanes-ANM-2013}, with an integration time step $dt = 0.06~t_u$, which results in relative errors of the total energy $Err(t)<10^{-7}$ for \textit{all} times. This particular integration scheme has been shown to perform very well in balancing computational speed and numerical accuracy for multidimensional Hamiltonian lattices \cite{ref-Danieli-MINE-2019}, and was successfully used for examining chaos in graphene \cite{ref-Hillebrand-Chaos-2020}.

The relaxed equilibrium positions that have previously been determined for the given value of stress $\sigma$ correspond to a graphene sheet being at zero temperature, without thermal fluctuations.
In order to simulate the system at finite temperatures, following a suitable energy-temperature calibration (see subsection \ref{section-E-T} below), we randomly insert an additional energy density (average energy per site) $e_N$ on the relaxed $T=0$~K state.
This additional energy is initially provided as solely potential energy, in the form of small random displacements of each atom from the relaxed zero temperature positions.
Then these displacements are properly scaled in order to adjust the added energy density $e_N$ to the desired value corresponding to the simulated temperature.
During evolution the initial potential energy gets shared into kinetic and potential energy and eventually the system equilibrates.

In general, for the numerical results presented in the next section we consider 10-20 different individual realizations of the randomly added initial energy, but we have selectively checked the robustness of the data when more realizations are used. For each realization we calculate the temporal evolution of the various quantities of interest, and then compute the average of these time-series over the different realizations in order to obtain the time dependence of the considered quantities for the ensemble. We denote such an averaged quantity over the different realizations with angled brackets, e.g. $\left\langle M(t)\right\rangle$ for the measurement of the quantity $M(t)$. We may further determine the average of a thermally equilibrated quantity over time. In such a case we average both over initial realizations and over time intervals, and we denote the computed average by using both an overline and angled brackets, e.g. $\overline{\left\langle M \right\rangle}$ for a variable $M$ at thermal equilibrium.

At finite temperatures the size of graphene sheets 
exhibits oscillations around their $T=0$~K relaxed configurations due to the thermal energy of the system (discussed further in Section~\ref{section-mechanical-response} below). In order to collect data over sufficiently many such sheet oscillations, we follow the system's time evolution up to $4\times10^3~t_u$. The recording window for all subsequent measurements is from $1\times10^3$ to $4\times10^3~t_u$, totaling $3000~t_u$. We have checked the insensitivity of the obtained results on the length of the recording window by testing longer time windows.

%%%%%%%%%%%%%%%%%%%%%%%%%%%%%%%%%%%%%%%%%%
\section{Results and Discussion}
\label{section-results}

%%%%%%%%%%%%%%%%%%%%%%%%%%%%%%%%%%%%%%%%%%
\subsection{Temperature calibration}
\label{section-E-T}

When an energy density $e_N$ is inserted in the strained graphene lattice, we observe that initially the total kinetic energy increases with time from its zero starting value and then following some relatively large fluctuations 
the system is settled at thermal equilibrium after at most $10^3~t_u$.
The time evolution of the system's temperature $T(t)$ towards equilibrium is computed in our microcanonical MD simulations through the energy equipartition relation
\begin{equation}
	\label{eqn-temperature}
	T(t) = \frac{E_K(t)}{N\,k_B}
	% T(t) = \frac{1}{k_B\cdot N_i\times N_j}\sum_{i,j}U(i,j)(t),
\end{equation} 
where $k_B$
%=8.617333262\times 10^{-5}$~eV/K  
is the Boltzmann constant.

In order to test whether thermal equilibrium has been reached, we compare the mean value and the standard deviation of the fluctuating temperature over various time windows.
Before achieving thermal equilibrium the standard deviation of the time-averaged $T$ is relatively large and also changes depending on the time window. When thermal equilibrium is reached the temperature fluctuations and the standard deviation are consistently small.
The mean temperature at thermal equilibrium is calculated by averaging over both the individual realizations and the recording time-window. We denote this average value $\overline{\left\langle T \right\rangle}$ by $T_{ave}$. In this case, the standard deviation of the measured values is computed using all data points over realizations and time.

\begin{figure}[H]
	\centering
	\includegraphics[width=7cm]{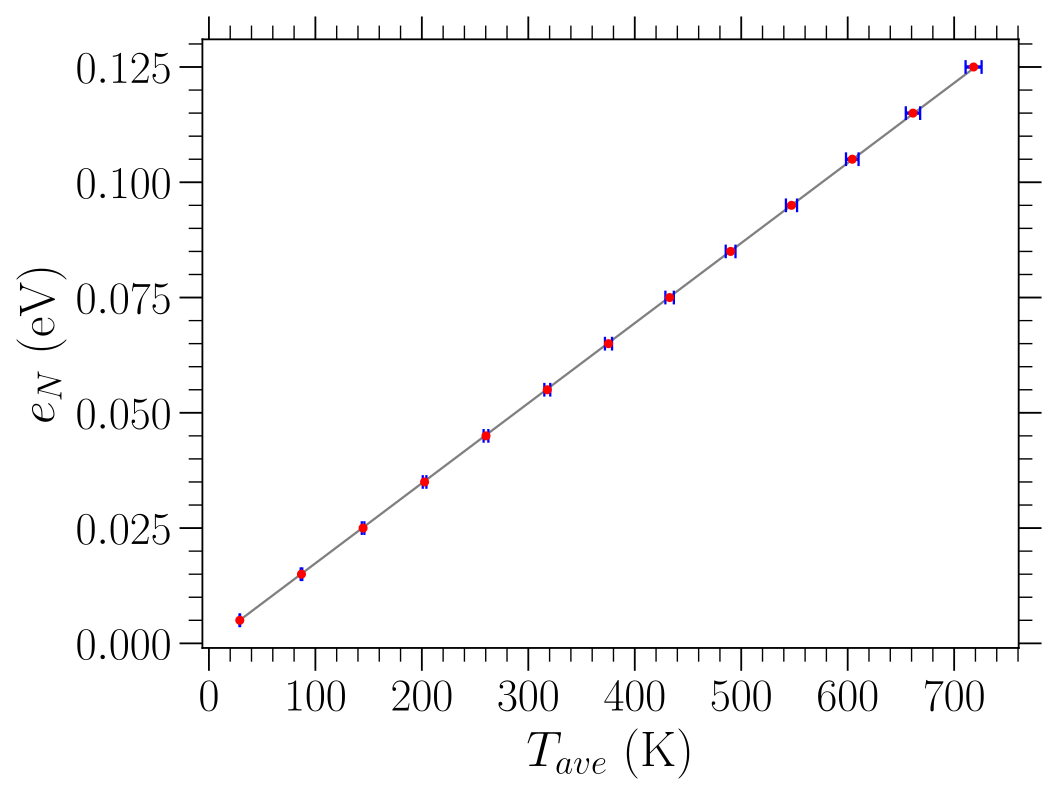}
	\caption{Symbols represent the relation between the additional energy density $e_N$ above the relaxed $T=0$~K graphene structure subjected to uniaxial tensile stress $\sigma=2.16$~eV/\AA$^2$ along the zigzag direction, and the average temperature $T_{ave}$ at thermal equilibrium, evaluated through the MD simulations by averaging over both time and the different realizations. One standard deviation of the $T_{ave}$ measurements is indicated by blue horizontal error-bars. The linear fitting of the presented data points is shown by gray solid line, providing a slope equal to $1.74\times10^{-4}$~eV/K.}
	\label{fig-energy-temperature}
\end{figure}

The relationship between the additional energy density $e_N$ on top of the relaxed equilibrium loaded structures and the averaged temperature $T_{ave}$ is linear in all cases of different stresses examined here, at least for temperatures up to 700~K considered in this work.
One representative case is shown in Figure \ref{fig-energy-temperature}.
The resulting slopes from the linear fittings of the data are very close for all values of stress $\sigma$ (a difference in the computed values is observed only in the fourth significant digit) and they are slightly above $2k_B$ due to the nonlinearities of the potential energy. For finite loads, the slope slightly increases with the amount of stress.
Thus, for a given value of stress $\sigma$, we use the obtained slope of the $e_N$ versus $T_{ave}$ linear fitting in order to control temperature (within a 1\% accuracy) in our investigation.
In particular, we  are setting the amount of the added energy density $e_N$ according to the desired temperature value.
%When we control the temperature in this way, we use the symbol $T$ to indicate the chosen temperature.

%============================
%============================

In order to investigate the system's elastic and structural properties, discussed in the following sections, we collect data from the central region of the lattice for avoiding potential edge effects and thus representing the behavior of bulk graphene. In particular, this central region sub-lattice has an analogous geometry to the larger structure, with number of columns $n_I= 44$ and number of rows $n_J= 45$.

\subsection{Mechanical response}
\label{section-mechanical-response}

% {(todo) mention in this section the robustness tests performed and precisely it affects the strain measurement (3rd/4th decimal changes)}

In the stress-controlled numerical implementation used here to examine the mechanical response of graphene, we compute the resulting strain due to the fixed force applied at the appropriate edge atoms. The uniaxial strain $\epsilon$ is obtained through the strain of the central row of the graphene sheet when the stress is applied along the armchair direction, while it is calculated by the average strain on the two central columns of the sheet in case of stress along the zigzag direction (see Figure~\ref{fig-lattice}). By $\epsilon_T$ is indicated the strain corresponding to temperature $T$.

In the zero temperature case, $T=0$~K, the uniaxial strain $\epsilon_0$ is determined through the relative change of the length of the central row (two central columns) of graphene subjected to a given applied stress along the armchair (zigzag) direction, with respect to the length of the central row (columns) at the unstrained equilibrium configuration shown in Figure~\ref{fig-lattice}. These measurements are taken in the central region of the lattice, as mentioned at the end of previous subsection.
For any length computations discussed here, a horizontal (vertical) length is measured as the difference of the $x$ ($y$) coordinates of the considered edge atoms.
The stress-strain response is obtained in this way at 0~K
% ...when we consider the length of a row, we measure this length purely in the horizontal direction (so the difference in $x$ coordinates of the last and first atoms of the bulk row for our setup in Figure~\ref{fig-lattice}), and the length of a column is measured purely in the vertical direction (so the difference in $y$ coordinates in our setup).

When the temperature of the system is raised at finite values, by adding energy to the equilibrated graphene, the lattice stretches and compresses in an oscillatory manner. The details of these oscillations depend on the temperature and the applied stress, and will be investigated in the future. 
In this case one has to take into account that the strain measurement $\epsilon_T(t)$ is now exhibiting temporal oscillations. Since we consider 10-20 different realizations of the randomly inserted initial energy distribution, we register the average, over these realizations, strain in time $\left\langle\epsilon_T(t)\right\rangle$, noting that the aforementioned oscillations are in-phase in the different realizations.

{For evaluating the strain $\epsilon_T$ of a uniaxially loaded graphene sheet at finite temperatures, a reference length {$\ell_T^{ref}$} corresponding to zero stress $\sigma=0$ at the particular value of $T$ is needed. This reference length accounts for thermal effects on the initial configuration and it is obtained by calculating the average, over realizations and time, of the length of the central row, or columns, of the sheet in the absence of any load. Then, when a stress is applied 
	the time evolution of strain in a particular realization is determined as the relative change of the length with respect to the reference length $\ell_T^{ref}$ }
\begin{equation}
	\label{eqn-strain}
	\epsilon_T(t) = \frac{\ell(t)-{\ell_T^{ref}}}{{\ell_T^{ref}}},
\end{equation} 
where $\ell(t)$ is the length at time $t$ of the central row or the average length of the two central columns depending on the direction of the applied uniaxial load.

% (length taken only in the horizontal direction of Figure \ref{fig-lattice})
% (lengths taken only in the vertical direction of Figure \ref{fig-lattice})

\begin{figure}[H]
	\centering
%	\begin{adjustwidth}{-\extralength}{0cm}
		\centering{\includegraphics[width=17cm]{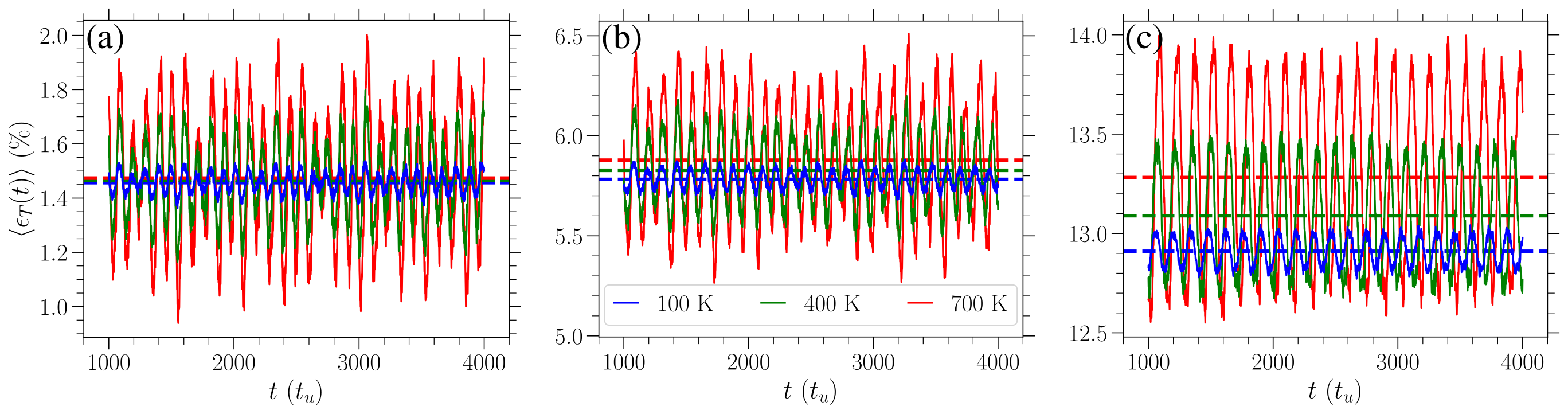}}
%	\end{adjustwidth}
	\caption{Time evolution of the average (over individual realizations) strain $\left\langle \epsilon_T(t)\right\rangle$, Equation~\eqref{eqn-strain}, when a stress \textbf{(a)} $\sigma=0.188$~eV/\AA$^2$, \textbf{(b)} $\sigma=1.03$~eV/\AA$^2$, and \textbf{(c)} $\sigma=1.97$~eV/\AA$^2$, along the zigzag direction is applied, for different temperatures: $T=100$~K (blue curves), $T=400$~K (green curves), and $T=700$~K (red curves). The average (over realizations and time) strains $\overline{\left\langle \epsilon_T\right\rangle}$, for each temperature, are indicated by the horizontal dashed lines of the same color in each panel. 
		\label{fig-strain_oscillations}}
\end{figure}

In Figure~\ref{fig-strain_oscillations} we highlight the behavior of $\left\langle{\epsilon_T}(t)\right\rangle$ for various values of stress $\sigma$ along the zigzag direction, at three distinct temperatures $T$ shown by different colors. An increase in temperature leads to an increase in the amplitude of strain oscillations as well as in an increase in the average strain. The latter one is obtained as the average over both realizations and time, $\overline{\left\langle{\epsilon_T}\right\rangle}$ and it is indicated by the dashed horizontal lines of different colors depending on the temperature in Figure~\ref{fig-strain_oscillations}. The average strains $\overline{\left\langle{\epsilon_{700}}\right\rangle}$ (red horizontal dashed lines) about which the $T=700$~K curves oscillate are higher than the $\overline{\left\langle{\epsilon_{100}}\right\rangle}$ (blue horizontal dashed lines) of the $T=100$~K curves in all cases of different stress. However, these differences are larger on absolute values for larger stresses. 
%============================

%============================

{We have checked that if one follows an alternative path on the $(\sigma,T)$-plane by giving first initial energy to the system and then applying forces at the edges, practically identical average strains are obtained. However, our approach is much more efficient because the temperature is accurately controlled and, more importantly, the system reaches thermal equilibrium significantly faster; in the alternative method the equilibration takes orders of magnitude longer.}

\begin{figure}[H]
	\centering
%	\begin{adjustwidth}{-\extralength}{0cm}
		\centering{\includegraphics[width=17cm]{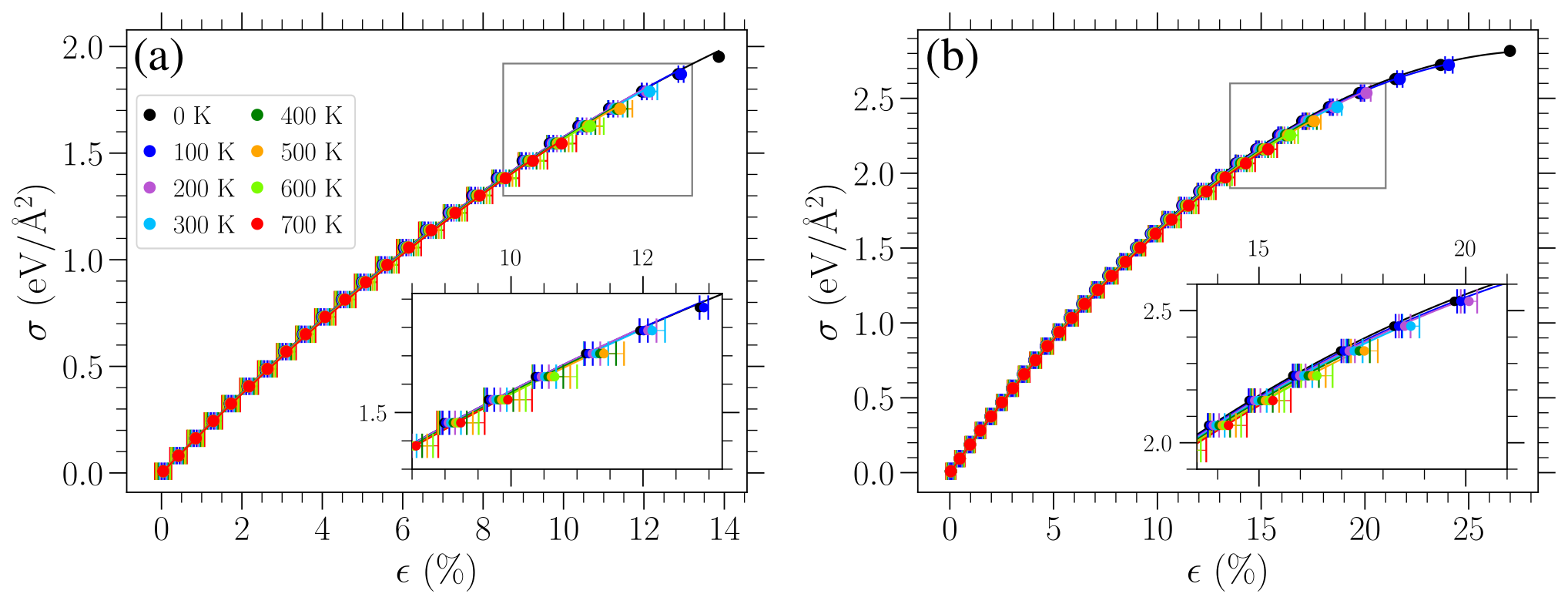}}
%	\end{adjustwidth}        
	\caption{Stress-strain response of planar graphene for uniaxial loads along the \textbf{(a)} armchair \textbf{(b)} zigzag direction, for different temperatures as indicated in the legend. Filled circles indicate the obtained average strain for each given stress. Solid curves represent fittings of these data with Equation~\eqref{eqn-stress-strain}, see text. For $T \ne 0$~K the strain is given as the average over time and realizations,  $\overline{\left\langle\epsilon_T\right\rangle}$, and the error-bars correspond to one standard deviation. {The insets in each panel depict a close-up view of the region indicated by the grey rectangle in each panel.}}
	
	\label{fig-stress-strain}
\end{figure}

Calculating the average strain $\overline{\left\langle{\epsilon_T}\right\rangle}$ as mentioned above, the mechanical response of planar graphene at different temperatures is obtained.
Stress-strain curves for uniaxial tensile loads along the armchair and zigzag directions are presented for various temperatures in Figure~\ref{fig-stress-strain}.
Despite the small differences, one can see for larger stresses that the average strain is a bit further to the right for the higher temperature cases. The error-bars indicate the standard deviation of the average strain measurement. As one can also deduce from Figure~\ref{fig-strain_oscillations}, the standard deviation is higher for higher temperatures. {This is highlighted via the insets in each panel, where a close-up of the data points and error-bars is presented for the region which is indicated by the grey rectangle in each panel. Close-up it is easier to see that the lengths of the error-bars increase with temperature.}

We have checked the accuracy of the presented strain measurements when more realizations or longer time windows are considered. In particular, the obtained strain values differ in the 3rd significant digit at most, when increasing the number of realizations or doubling the length of the time window.
%============================
%============================

Since at finite temperatures the strain is measured with respect to {the averaged oscillating length $\ell_T^{ref}$ due to thermally induced vibrations of the unstrained sheet,} the stress-strain curves pass from the origin of Figure~\ref{fig-stress-strain}, as expected. From Figure~\ref{fig-stress-strain}, we see that the temperature has a relatively small effect on the stress-strain response, at least for the values of $T$ considered here,
apart from the significant reduction of the fracture point. For small stresses the achieved strain is practically the same for the two directions of applied stress, while the strong directional dependence at large stresses has already been well established in previous investigations \cite{ref-Liu-PRB-2007,ref-Zhao-NanoLet-2009,ref-Kalosakas-JAP-2013,ref-deSousa-Intech-2021}. 

\begin{figure}[H]
	\centering
	\centering{\includegraphics[width=14cm]{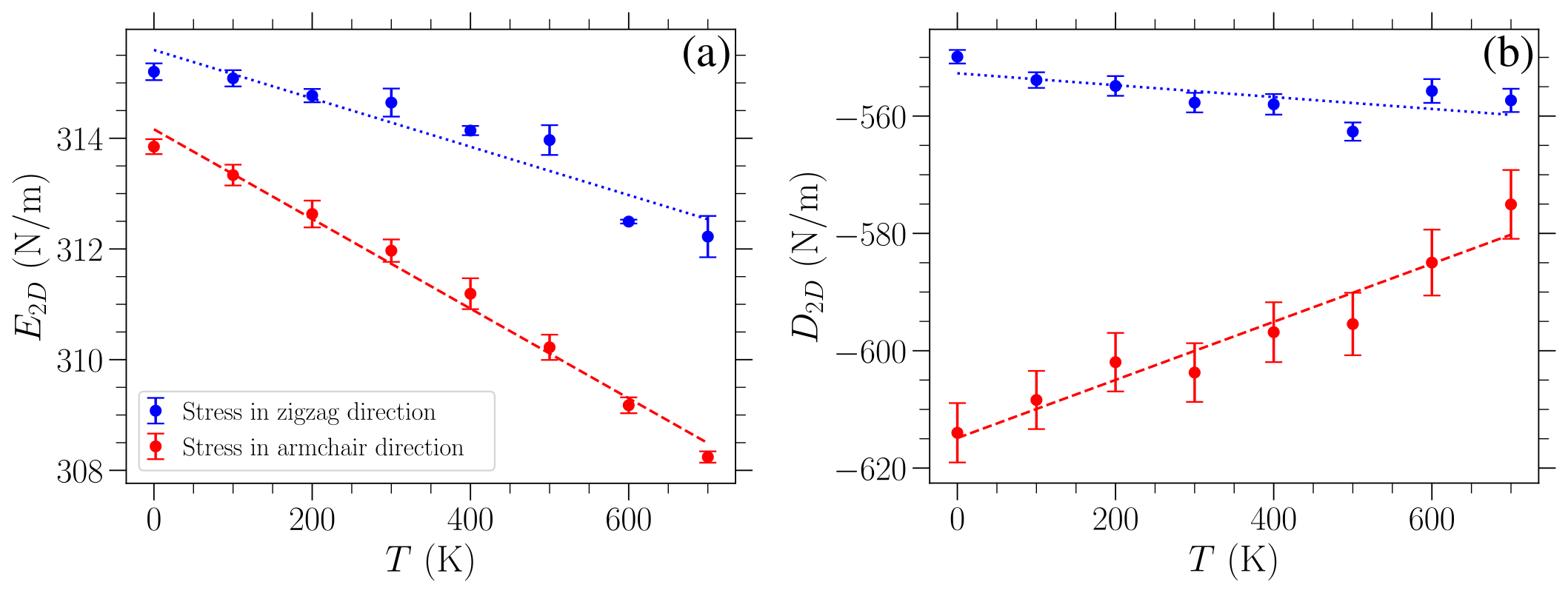}}
	\caption{Temperature dependence of \textbf{(a)} the {Young} modulus $E_{2D}$ and \textbf{(b)} the third-order elastic modulus $D_{2D}$, for applied stress along the armchair (red points) or the zigzag (blue points) direction, evaluated through fittings of the data of Figure~\ref{fig-stress-strain} with Equation~\eqref{eqn-stress-strain} (see text). }
	\label{fig-moduli}
\end{figure}

The stress-strain response can be described by the nonlinear relation~\cite{ref-Kalosakas-JAP-2013}
\begin{equation} 
	\sigma = E_{2D}\cdot\epsilon + D_{2D}\cdot\epsilon^2,
	\label{eqn-stress-strain}
\end{equation} 
where $\sigma$ is the applied uniaxial stress, $\epsilon$ the corresponding strain, $E_{2D}$ is the 2D {Young} modulus and $D_{2D}$ is the 2D third-order elastic modulus. For each temperature examined and both directions of applied stress, we first obtained the value of Young modulus by the linear response at small stress/strain and then we fit the data presented in Figure~\ref{fig-stress-strain} with Equation~\eqref{eqn-stress-strain} to determine the third-order elastic modulus.  
The computed values of $E_{2D}$ and $D_{2D}$ are plotted in Figure~\ref{fig-moduli} as a function of temperature, for applied stress in either the armchair (red points) or the zigzag (blue points) direction.
The error-bars on these points indicate one standard deviation of the fitted parameters under the observed covariance of the fit. A linear variation can roughly approximate the obtained temperature dependence of these elastic moduli. Linear fittings of the corresponding data are indicated by the dashed red (dotted blue) line for stress in the armchair (zigzag) direction. 
%============================

%============================

The {Young} modulus $E_{2D}$ [Figure~\ref{fig-moduli}(a)] appears to decrease almost linearly with increasing temperature, albeit only by a relatively small amount, which is consistent with other results in the literature \cite{ref-Shao-JCP-2012,ref-Shen-MatDes-2010,ref-Tahani-AMM-2024}. In particular, the linear fitting of these data for stress along the armchair direction leads to a variation of $E_{2D}(T)$ with a slope $- 8.1\times 10^{-3}$ (N/m)/K, 
%is \begin{equation}
	%    \label{eqn-EvsT-ac}
	%   E_{2D} = 314 - 8.1\times 10^{-3}~T,
	%\end{equation} 
	while for stress along the zigzag direction the slope is $ - 4.4\times 10^{-3}$ (N/m)/K.
	%\begin{equation}
	%    \label{eqn-EvsT-zz}
	%   E_{2D} = 316 - 4.37\times 10^{-3}~T.
	%\end{equation} 
	
	The decrease of {Young} modulus with temperature is often given in the literature as a percentage change over the investigated range of temperatures. {To enhance a direct quantitative comparison, we present our results along with existing ones in the literature as \textit{percentage change per} 100~K. In our case the {Young} modulus decreases by $0.25\%/100$~K and $0.13\%/100$~K for stress along the armchair or the zigzag direction, respectively. 
		Combined density functional theory and quasi-harmonic approximation calculations in Reference~\cite{ref-Shao-JCP-2012} have determined a $E_{2D}$ decrease by $0.22\%/100$~K over the range from 0~K to 1000~K.
		Molecular dynamics has been used in Reference~\cite{ref-Shen-MatDes-2010} for investigations at 300~K, 500~K, and 700~K of graphene lattices of different aspect ratios  consisting of 1886 atoms. $E_{2D}$ is found to decrease by $1.3\%/100$~K, $0.48\%/100$~K, and $0.33\%/100$~K (decrease by $0.88\%/100$~K, $0.65\%/100$~K, and $0.33\%/100$~K) for strain along the armchair (zigzag) direction, where the three different values correspond to graphene aspect ratios 1.97, 1.44, and 1.01 (1.95, 1.45, and 0.99). We note that our lattice has an aspect ratio 1.72.}
	{Molecular dynamics simulations in Reference \cite{ref-Zhang-CompMatSci-2013} have resulted in a $E_{2D}$ decrease by $1.4\%/100$~K over a temperature range of 300~K to 2000~K, for strain applied along the armchair direction.}
	{Finally, in Reference~\cite{ref-Tahani-AMM-2024} a reduction in $E_{2D}$ between $0.19\%/100$~K and $0.25\%/100$~K has been obtained for temperatures ranging from 0~K to 1600~K, where the varying reduction depends on the different parameterizations of the used model, which affects the Young modulus value at $T=0$~K.}
	
	{The calculated Young modulus $E_{2D}$ at 300~K is in good agreement with values reported in experimental studies conducted at room temperature \cite{ref-Lee-Science-2008,ref-Cao-NatComm-2020,ref-Varillas-MechSci-2024}. In particular, our findings, namely $315~$N/m and $312~$N/m for loading along the zigzag and armchair direction, respectively, are consistent with the Young modulus reported in References \cite{ref-Lee-Science-2008} ($340 \pm 50~$N/m), \cite{ref-Cao-NatComm-2020} ($300$ to $340~$N/m), and \cite{ref-Varillas-MechSci-2024} ($350 \pm 100~$N/m).}
	
	We observe from Figure~\ref{fig-moduli}(b) that the $D_{2D}$ values are consistently higher for strain in the zigzag direction, than for the other direction.
	This is congruent with the fact that the graphene sheet 
	is more resistant to stress along the zigzag direction.
	When stress is along the armchair direction, one third of all the bonds are parallel to the direction of strain, and hence these bonds exhibit maximal stretching in the sheet.
	Taking into account the respecting angle deformations there is in general a higher strain for the same stress in this case as compared to loads along the zigzag direction. Considering that the Young modulus is almost the same in this two cases, this leads to lower $D_{2D}$ modulus (i.e. higher absolute values) for stresses in the armchair direction. The different strains for a given stress in the two perpendicular loading directions discussed here can be seen when comparing the panels of Figure~\ref{fig-stress-strain}, where the curves for stress in the armchair direction lie further to the right than when the stress is applied in the zigzag direction, i.e., indicating higher strains for the same stress.
	Regarding the temperature variation of $D_{2D}$, different trends are exhibited when the stress is along the zigzag or the armchair direction.
	A linear fitting of the $D_{2D}(T)$ data points
	results in a slope $+5.0\times 10^{-2}$ (N/m)/K
	for strain in the armchair direction
	%\begin{equation}
	%   \label{eqn-DvsT-ac}
	%  D_{2D} = -615 + 4.95\times 10^{-2}~T,
	%\end{equation}
	and $-1.0 \times 10^{-2}$ (N/m)/K
	for strain in the zigzag direction. The value of $D_{2D}$ increases by $0.91\%/100$~K (decreases by $0.19\%/100$~K) for strain in the armchair (zigzag) direction over the temperature range from 0~K to 700~K. 
	
	%\begin{equation}
	%    \label{eqn-DvsT-zz}
	%    D_{2D} = -553 - 1.01\times 10^{-2}~T.
	%\end{equation}
	
	% from output:
	% strain in armchair direction: we have 
	% E = -0.008101471908555036 * T + 314.16250415419097
	% strain in zigzag direction: we have 
	% E = -0.00436982018027976 * T + 315.5950043328857
	% strain in armchair direction: we have 
	% D = 0.04951992492233959 * T + -614.8772026984076
	% strain in zigzag direction: we have 
	% D = -0.01011885844461724 * T + -552.7133797451606
	
\begin{figure}[H]
		\centering
		\includegraphics[width=14cm]{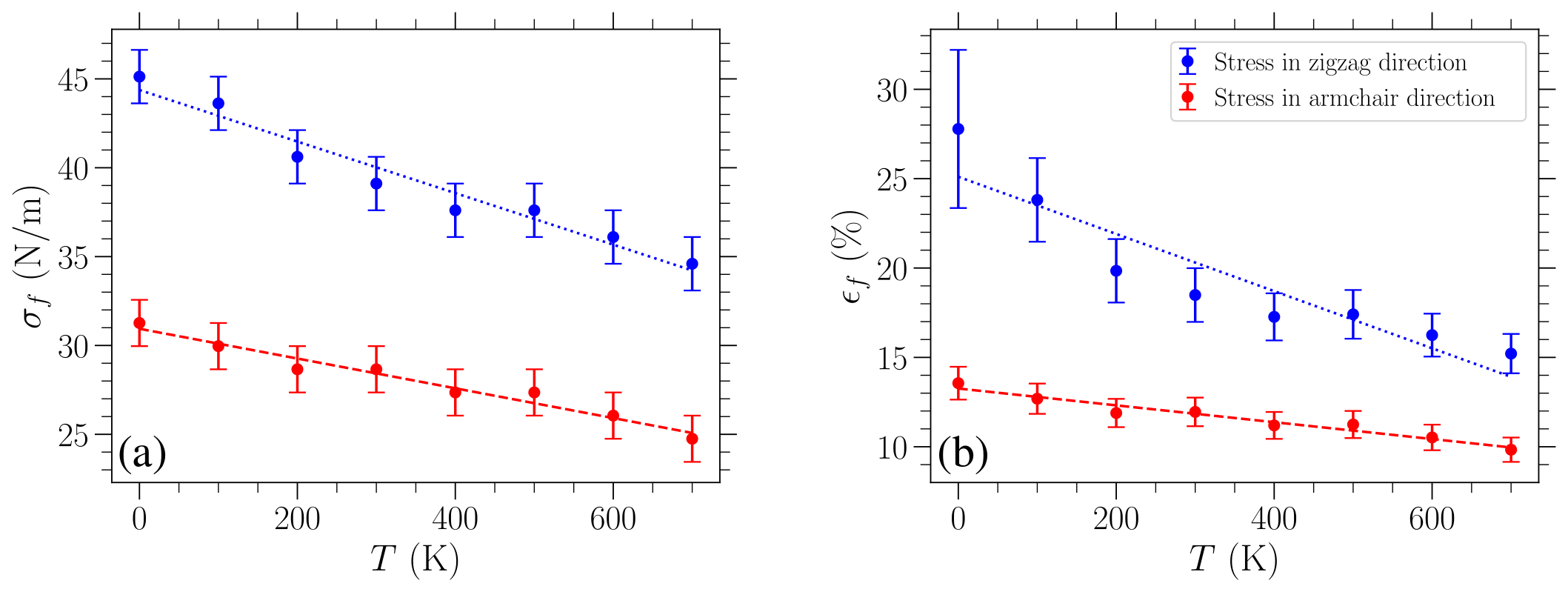}
		\caption{ Temperature dependence of (a) the fracture strength $\sigma_{f}$ and (b) the associated failure strain $\epsilon_{f}$ of graphene. Straight lines represent linear fittings.}
		\label{fig-fracture}
	\end{figure}

	Finally, we estimate the graphene's fracture strength $\sigma_{f}$ and failure strain $\epsilon_{f}$, for different temperatures $T$. The former one is obtained by the highest tested value of stress $\sigma$ which does not lead to failure of the graphene sheet. Its error-bar is provided by the step we use in the increment of the tested $\sigma$ values, which are evenly spaced. These results are presented in Figure~\ref{fig-fracture}(a),
	where an almost linear decrease of the fracture stress with temperature is shown.
	A linear fitting of these data points is indicated with a dashed red (dotted blue) line for stress in the armchair (zigzag) direction. The slope of the linear fitting of the $\sigma_f(T)$ data is $-8.4\times10^{-3}$~(N/m)/K for stress in the armchair direction and $-1.5\times10^{-2}$~(N/m)/K for stress in the zigzag direction.
	
	Such a linear dependence of the fracture strength on temperature is in accordance with existing results. {In particular, we estimate that, in the MD study of graphene loaded along the armchair direction reported in Reference~\cite{ref-Zhao-JAP-2010}, the fracture strength decreases with temperature at a rate of $-8.6\times 10^{-3}$~(N/m)/K, which is in very good agreement with the slope obtained in our work for the same direction. Molecular dynamics simulations are also used in Reference~\cite{ref-Tanhadoust-DiamRelMat-2022} and the presented temperature dependence of the fracture stress decreases linearly with a slope of $-1.4\times 10^{-2}$~(N/m)/K, however it is not clear which was the loading direction there.
		%but based on the stated zero-temperature fracture stress of 131.1~GPa (about 44~N/m) as well as the stress-strain curves presented for both directions in Figure 12 of this paper, it is reasonable to assume that the presented data (Figure 4 of the work) is for the zigzag direction.
		In another MD investigation, where the stated loading direction was the armchair direction, the corresponding slope was found $-1.6\times10^{-2}$~(N/m)/K~\cite{ref-Zhang-CompMatSci-2013}. Lastly, Reference~\cite{ref-Varma-IntJMechMatDes-2022} employs a combination of machine learning and MD simulations with the Tersoff potential. In that study, the fracture strength decreases linearly with temperature, exhibiting a slope of $-3.8\times10^{-2}$~(N/m)/K for loading along the zigzag direction, whereas for loading along the armchair direction a bilinear behavior is observed: the slope is initially $-6.7\times10^{-3}$~(N/m)/K but becomes steeper above 500~K, thus representing the only non-linear trend reported in the literature.}

	%{At $T=0$~K our intrinsic strength is $135$~GPa ($93$~GPa) for stress in the zigzag (armchair) direction, assuming an effective thickness of 0.335~nm of the graphene sheet, which is in alignment with the experimentally determined intrinsic strength $130\pm10$~GPa \cite{ref-Lee-Science-2008}.}
	%============================
	
	The failure strain, $\epsilon_{f}$, at different temperatures has been estimated through the value of fracture stress by solving for $\epsilon_{f}$ in Equation~\eqref{eqn-stress-strain}. To this end, the known value of $\sigma_{f}$ as well as the fitted values of $E_{2D}$ and $D_{2D}$ describing the stress-strain curve at the given temperature, have been used.  In this case the error-bars are determined by converting the corresponding extreme values of stress, $\sigma_{f} \pm \Delta \sigma_{f}$,
	to strain [via Equation~\eqref{eqn-stress-strain}], and then choosing the maximum absolute difference from $\epsilon_{f}$. These results are shown in Figure~\ref{fig-fracture}(b), where again a linear fitting of the data is indicated with a dashed red (dotted blue) line for stress in the armchair (zigzag) direction. 
	The linear fitting of the $\epsilon_f(T)$ data points leads to a slope $ - 4.7\times10^{-3}$~\%~strain/K for stress along the armchair direction and $- 1.6\times10^{-2}$~\%~strain/K for stress in the zigzag direction. {We can further compare these slopes to existing results by estimating the slopes of the failure strain versus temperature data reported in the literature. For uniaxial loading along the armchair direction in Reference~\cite{ref-Zhao-JAP-2010}, a slope of $-4.8\times 10^{-3}$~\%~strain/K is estimated, which is in very good agreement with our result. A slope of $-5.4\times 10^{-3}$~\%~strain/K is estimated for the results presented in Reference~\cite{ref-Tanhadoust-DiamRelMat-2022}, where, as noted above, it is not clear which is the loading direction. Finally, the corresponding slope for strain applied along the armchair direction is $-5.9\times 10^{-3}$~\%~strain/K in Reference~\cite{ref-Zhang-CompMatSci-2013}.}
	
	{The fracture stress reported in the experimental study of Reference \cite{ref-Lee-Science-2008} is $42 \pm 4$~N/m. We have obtained an intrinsic strength of $39.1 \pm 1.5~$~N/m ($28.7 \pm 1.3$~N/m) and a fracture strain of $18.5 \pm 1.5\%$ ($12.0 \pm 0.8\%$) for loading along the zigzag (armchair) direction at 300~K. We note that our results are in good agreement with the intrinsic strength and fracture strain values reported in other studies which additionally allow out-of-plane deformations of the material. Specifically, MD simulations carried out at 300~K using the AIREBO potential, yield, for loading along the zigzag direction, intrinsic strengths around $36-37$~N/m and fracture strains between $17-20\%$ \cite{ref-Zhao-NanoLet-2009,ref-Varillas-MechSci-2024}. 
		For loading along the armchair direction,
		Reference \cite{ref-Lee-NanoResLets-2015} conducted MD simulations using the REBO force field at 300~K, obtaining a fracture strain of $12.5\%$ at a stress of $29.1$~N/m, while Reference \cite{ref-Zhao-NanoLet-2009} using AIREBO reports an intrinsic strength of $30~$N/m and fracture strain $13\%$. }
	
	{The results shown in Figure~\ref{fig-fracture} indicate that graphene fails at lower applied stress/strain as temperature increases.
		This is reasonable since, as can be clearly seen from Figure~\ref{fig-strain_oscillations}, for the stress-controlled simulations considered here the sheet achieves higher strains over the course of its oscillations by increasing temperature. As a result, for fixed stress the bonds between neighboring atoms experience longer stretches at higher temperatures and therefore are more likely to break, causing failure of the material, due to the increase of the maximum deformation of the lattice. 
		Moreover, graphene can tolerate higher loads along the zigzag direction in the whole temperature range investigated here, as is implied by the results of Figure~\ref{fig-fracture}, where the values of fracture strength and failure strain are consistently lower for stress in the armchair direction (red data points) as compared to stress applied in the zigzag direction (blue data points).
		In contrast to the relatively stronger temperature dependence of the fracture strength and failure strain, the Young modulus variation shown in Figure~\ref{fig-moduli}(a) exhibits a much smaller relative change, implying that the influence of thermal effects on the stiffness of graphene is less significant, at least within this temperature regime. }

	%%%%%%%%%%%%%%%%%%%%%%%%%%%%%%%%%%%%%%%%%%
	\subsection{Bond length and bond angle distributions}
	\label{section-distributions}
	
	In order to analyze the effects of temperature and stress on the distributions of the lengths and angles of the bonds, we first distinguish the two types of bond lengths, denoted by $A$ and $Z$, and the two types of angles, indicated by $\alpha$ and $\zeta$, as illustrated in Figure~\ref{fig-lattice}. The $A$ bonds are along the armchair direction.
	The $Z$ bonds alternate symmetrically along the zigzag direction and both exhibit identical deformations at 0~K when a uniaxial stress is applied along the high symmetry zigzag or armchair directions.
	The angles $\alpha$ and $\zeta$ represent the bond angles formed between two consecutive $Z$ bonds and between an $A$ and a $Z$ bond, respectively. They respond always oppositively under an applied stress, due to the geometry of the system and the constraint $\alpha +2\zeta=2\pi$. 
	
	When a load is applied at zero temperature, due to the absence of fluctuations and the static nature of the strained sheet, there is no variability in the two types of bond lengths and angles and their distribution is delta-peaked. Approximate expressions for the strain dependence of bond lengths $A$ and $Z$ and angles $\alpha$ and $\zeta$ were provided in Reference~\cite{ref-Kalosakas-Mat-2022}. Indicating by the indices $a$ or $z$ a load applied along the armchair or zigzag direction, respectively, these expressions read
	\begin{align}
		\label{eqn-ac-lengths-A}
		A_a&= 1.42 + 0.011\,\epsilon_{0}+0.00024\,\epsilon^2_{0},\\ 
		Z_a&= 1.42+0.0031\,\epsilon_{0}-0.000046\,\epsilon_{0}^2,
		\label{eqn-ac-lengths-Z}
	\end{align}
	% the function $\epsilon_{zz}(\Sigma) = \frac{-320/1602 + \sqrt{(320/1602)^2+4(-560/160200)\Sigma}}{2\cdot(-560/160200)}$ 
	% has been derived from the stress-strain relation at zero temperature. 
	and
	\begin{align}
		\label{eqn-ac-angles-alpha}
		\alpha_a &= 120^\circ -0.83\,\epsilon_0+0.020\,\epsilon^2_0,\\
		\zeta_a &= 120^\circ + 0.41\,\epsilon_0 - 0.010\,\epsilon^2_0,
		\label{eqn-ac-angles-zeta}
	\end{align} 
	while for stress along the zigzag direction
	\begin{align}
		\label{eqn-zz-lengths-A}
		A_z&= 1.42,\\
		Z_z&=1.42+0.0088\,\epsilon_0+0.000080\,\epsilon_0^2,
		\label{eqn-zz-lengths-Z}
	\end{align} 
	and
	\begin{align}
		\label{eqn-zz-angles-alpha}
		\alpha_z &= 120^\circ+ 0.80\,\epsilon_0-0.013\,\epsilon^2_0\\
		\zeta_z &= 120^\circ -0.40\,\epsilon_0+0.0064\,\epsilon^2_0.
		\label{eqn-zz-angles-zeta}
	\end{align} 
	In Equations \eqref{eqn-ac-lengths-A}, \eqref{eqn-ac-lengths-Z}, \eqref{eqn-zz-lengths-A} and \eqref{eqn-zz-lengths-Z}, the bond lengths $A$ or $Z$ are given in \AA~and the zero temperature strain $\epsilon_0$ is expressed as \%~strain. Similarly, in Equations \eqref{eqn-ac-angles-alpha}, \eqref{eqn-ac-angles-zeta}, \eqref{eqn-zz-angles-alpha} and \eqref{eqn-zz-angles-zeta} the bond angles $\alpha$ and $\zeta$ are provided in degrees and $\epsilon_0$ should be given again in percentage strain. 
	The distributions of bond lengths (angles) in bulk graphene at $T=0$~K are given by double singular peaks at the locations provided by the above pairs of relations for the bond lengths (angles), depending on the direction of the loading, for different values of the applied uniaxial strain. 
	
	In order to reveal the influence of temperature on the bond lengths and angles distributions, we register all the fluctuating bond length and angle values during the system's evolution in our measurement window and obtain normalized distributions for different amounts of stress/stain at various temperatures. In particular, we create a distribution for each realization by allocating all the measured bond lengths (angles) into fine-grained bins of width $3.5\times10^{-3}$~\AA ~($0.004\,\frac{180^\circ}{\pi}$). The resulting distributions are normalized and then averaged over the different realizations in order to obtain the final distribution for each case. It is worth noting that the size of the error-bars, indicating one standard deviation of this averaging computation over the different realizations, are negligible, and hence not included in the plots of the distributions presented below.
	%in Figures~\ref{fig-lengths},~\ref{fig-angles},~\ref{fig-zero_stress_distros_variances}(a) - (b),~\ref{fig-lengths_fitted},~and~\ref{fig-angles_fitted}. 
	We emphasize again that we consider the central region of the sheet for collecting our data, as mentioned at the end of Section~\ref{section-E-T}.
	
\begin{figure}[H]
		\centering
%		\begin{adjustwidth}{-\extralength}{0cm}
			\centering{\includegraphics[width=17cm]{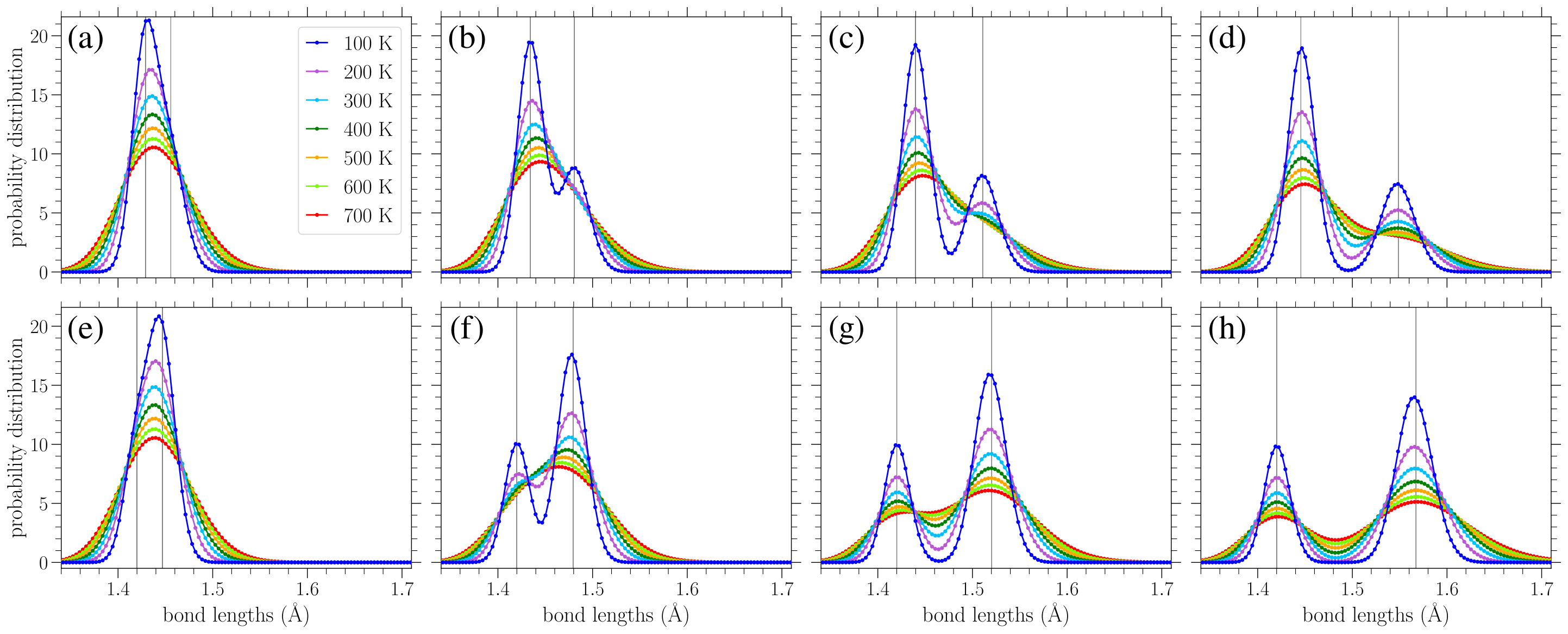}}
%		\end{adjustwidth}
		\caption{Normalized bond length distributions in graphene, for increasing applied stress (left to right), along the armchair (top row) and the zigzag (bottom row) direction, at different temperatures $T$ as indicated in the legend. The curves are guides to the eyes. The vertical lines indicate the values of the $A$ and $Z$ bond lengths at zero temperature, from Equations \eqref{eqn-ac-lengths-A}, \eqref{eqn-ac-lengths-Z}, \eqref{eqn-zz-lengths-A}, and \eqref{eqn-zz-lengths-Z}. The stresses $\sigma$ in the armchair direction are \textbf{(a)} 0.569~eV/\AA$^2$, \textbf{(b)} 0.895~eV/\AA$^2$, \textbf{(c)} 1.22~eV/\AA$^2$, and \textbf{(d)} 1.55~eV/\AA$^2$, while in the zigzag direction are \textbf{(e)} 0.563~eV/\AA$^2$, \textbf{(f)} 1.13~eV/\AA$^2$, \textbf{(g)} 1.69~eV/\AA$^2$, and \textbf{(h)} 2.16~eV/\AA$^2$.}
		\label{fig-lengths}
	\end{figure}

	In Figures~\ref{fig-lengths}(a) - (d) we show the normalized bond length distributions for increasing values of stress applied along the armchair direction. In Figures~\ref{fig-lengths}(e) - (h), the applied load is along the zigzag direction and the stress increases from (e) to (h) too. 
	%We note that all plots are on the same set of axes.
	When there is no loading, $\sigma=0$, at finite temperatures the distributions are simply normal distributions with the variance linearly increasing with temperature (see Figure~\ref{fig-zero_stress_distros_variances}(a) below). Increased temperature leads to larger fluctuations in the lattice, resulting in a wider spread of the observed bond lengths. 
	In the presence of uniaxial loading, for the smaller values of stress presented in Figures~\ref{fig-lengths}(a) and (e), there is a slight skewing of the distributions. As the stress is increased the single peak splits into two peaks which are gradually separated more and more as can be clearly seen from the plots corresponding to lower temperatures, due to the increased separation of the $A$ and $Z$ bond length values (see Figures 2 and 3 of Reference~\cite{ref-Kalosakas-Mat-2022}). However, the increase of temperature leads to the merging of these two peaks due to their broadening. The centers of the peaks correspond to the zero-temperature values of the two types of bond lengths for each different direction of the applied stress as given in Equations \eqref{eqn-ac-lengths-A} - \eqref{eqn-ac-lengths-Z} and \eqref{eqn-zz-lengths-A} - \eqref{eqn-zz-lengths-Z}.
	%============================
	%============================
	
	Since there are twice as many $Z$ type bond lengths as $A$ types, the highest peak in each distribution in Figure~\ref{fig-lengths} is mostly encompassing the lengths of the $Z$ type bonds. Thus, we can see that for stress applied along the armchair direction [Figures~\ref{fig-lengths}(a) - (d)], it is the $A$ bonds which achieve greater lengths (the lower peak, further to the right) while the $Z$ bonds exhibit a smaller extension. In contrast, for stress applied along the zigzag direction [Figures~\ref{fig-lengths}(e) - (h)], the $Z$ bonds achieve greater lengths (the taller peak is to the right in the distributions) while the centers of the smaller peaks remain near $r_0=1.42$~\AA, in accordance with Equation~\eqref{eqn-zz-lengths-A} and the corresponding broadening due to thermal effects. The fact that all bonds stretch for stress applied along the armchair direction, but only the $Z$ type bonds (two-thirds of all the considered bonds) are extended for a load along the other direction~\cite{ref-Kalosakas-Mat-2022}, justifies that the gap between the two peaks is more pronounced for stress applied in the zigzag direction.
	
\begin{figure}[H]
		\centering
%		\begin{adjustwidth}{-\extralength}{0cm}
			\centering{\includegraphics[width=17cm]{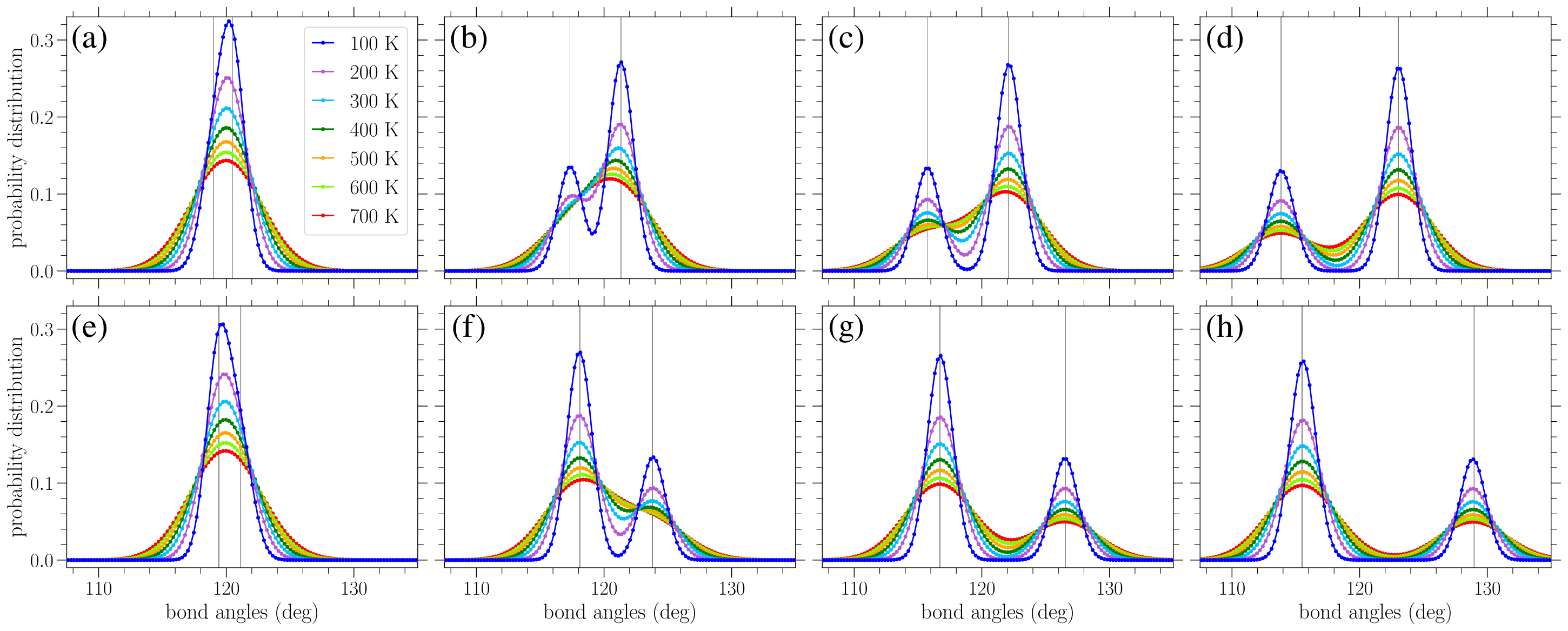}}
%		\end{adjustwidth}
		\caption{Normalized bond angle distributions in planar graphene, for increasing stress (left to right), applied along the armchair (top row) and in the zigzag (bottom row) direction, at various temperatures $T$ as indicated in the legend. {The curves are guides to the eyes. The vertical lines indicate the values of the $\alpha$ and $\zeta$ bond angles at zero temperature, from Equations \eqref{eqn-ac-angles-alpha}, \eqref{eqn-ac-angles-zeta}, \eqref{eqn-zz-angles-alpha}, and \eqref{eqn-zz-angles-zeta}.} The values of stress $\sigma$ along the armchair direction are \textbf{(a)} 0.244~eV/\AA$^2$, \textbf{(b)} 0.651~eV/\AA$^2$, \textbf{(c)} 1.06~eV/\AA$^2$, and \textbf{(d)} 1.55~eV/\AA$^2$, while in the zigzag direction are \textbf{(e)} 0.282~eV/\AA$^2$, \textbf{(f)} 0.939~eV/\AA$^2$, \textbf{(g)} 1.60~eV/\AA$^2$, and \textbf{(h)} 2.16~eV/\AA$^2$.}
		\label{fig-angles}
	\end{figure}

	In Figure~\ref{fig-angles} similar results are presented as in Figure~\ref{fig-lengths}, but for the distribution of bond angles. At zero strain, normal distributions centered about the equilibrium angle of $\phi_0=120^\circ$ are obtained for finite temperatures, with a variance increasing with temperature (see Figure~\ref{fig-zero_stress_distros_variances}(b) below). Again we see the gradual peak splitting due to increased stress, while increasing the temperature leads to the broadening and merging of these peaks. The highest peak in the bond angle distributions corresponds to the $\zeta$ type angles, since there are twice as many $\zeta$ angles as $\alpha$ angles. For stress along the armchair direction [Figures \ref{fig-angles}(a) - (d)], the $\alpha$ angles decrease, while the $\zeta$ angles increase. Reverse is the situation when the stress is applied along the zigzag direction [Figures \ref{fig-angles}(e) - (h)]. Also in this case the peaks of the distributions are centered about the zero temperature $\alpha$ and $\zeta$ values, as given in Equations \eqref{eqn-ac-angles-alpha} - \eqref{eqn-ac-angles-zeta} or \eqref{eqn-zz-angles-alpha} - \eqref{eqn-zz-angles-zeta}, depending on the direction of the applied stress. 
	%============================
	%============================

	%%%%%%%%%%%%%%%%%%%%%%%%%%%%%%%%%%%%%%%%%%
	\subsubsection{Analytical expressions for the bond length and bond angle distributions}
	\label{section-fittings}
	
	We now present approximate analytical expressions for the bond length and angle distributions, as those shown in Figures~\ref{fig-lengths} and \ref{fig-angles}, in order to describe the dependence of graphene's structural properties on stress and temperature. Based on the results discussed in the previous subsection, we note that the obtained distributions appear to approximately be given through the combination of two normal distributions, where the means of these normal distributions correspond to the values of the two types of bond lengths, or angles, found for each stress at zero temperature. The variance of these normal distributions is induced by thermal fluctuations, while the difference in peak heights is related to the fact that there exist double as many of one type of bond length (or angle) as the other. 
	
	As there exist approximate expressions available for the equilibrium bond lengths and bond angles as a function of the applied strain at $T=0$~K [see Equations~\eqref{eqn-ac-lengths-A} - \eqref{eqn-zz-angles-zeta}], it remains to determine the explicit dependence of variance on temperature. This will be obtained by numerically evaluating the effects of temperature on the normal distributions of the bond lengths and bond angles, at the unstrained graphene sheet. The results of these calculations are compared with analytical estimates of the variance through the Boltzmann distribution,
	using a second-order approximation on the relevant potential energy terms describing bond stretching and angle bending.
	
\begin{figure}[H]
		\centering
		\includegraphics[width=14cm]{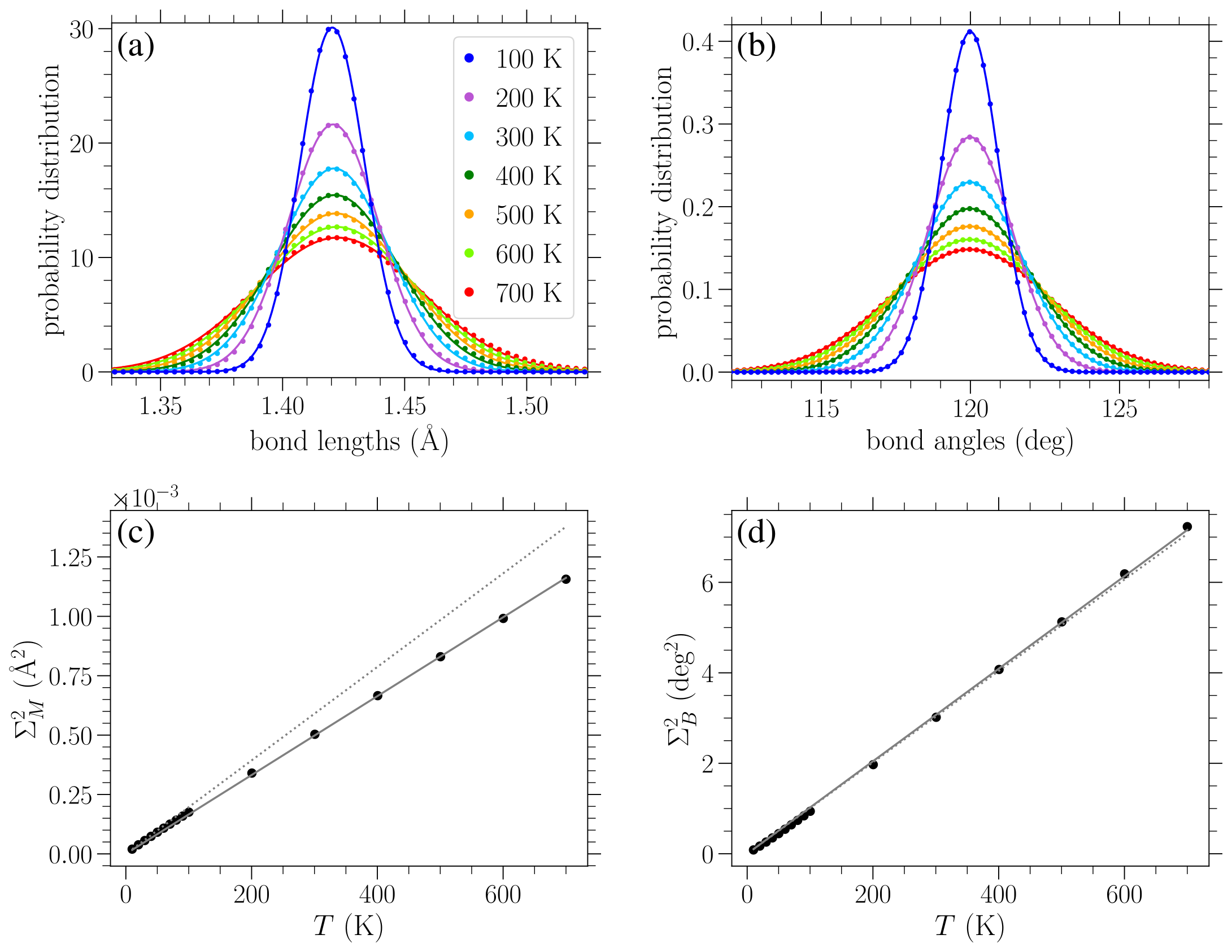}
		\caption{Normalized distributions of (\textbf{a}) bond lengths and (\textbf{b}) bond angles in bulk unstrained graphene ($\sigma=0$), at different temperatures $T$, as indicated in the legend, are shown by symbols. 
			Solid lines in (a) and (b) represent Gaussian fittings of the numerical data.
			Circles present the temperature dependence of \textbf{(c)} the  variance $\Sigma_M^2$ of the Gaussian fitting of the bond length distributions presented in (a) and \textbf{(d)} the variance $\Sigma_B^2$ of the Gaussian fitting of the bond angle distributions shown in (b). Solid lines in (c) and (d) indicate linear fittings of the corresponding data, see Equations \eqref{eqn-CM} and \eqref{eqn-CB}, while dotted lines denote the analytical approximating expressions of Equations \eqref{eq:S_M} and \eqref{eq:S_B}, respectively.}
		\label{fig-zero_stress_distros_variances}
	\end{figure}

	Performing a Gaussian curve fitting procedure to the numerically obtained distributions of the bond lengths and bond angles at zero applied stress for various temperatures, shown in Figures~\ref{fig-zero_stress_distros_variances}(a) and (b), respectively, we compute the corresponding variances and mean values. The dependence of these variances on temperature are presented by filled circles in Figures~\ref{fig-zero_stress_distros_variances}(c) and (d) for the bond length and angle distributions, respectively.
	Solid lines in the latter plots denote a linear fitting of the data.
	It is worth noting that the mean of the bond length distribution slightly increases with temperature too, due to the soft Morse potential describing bond stretching. However, incorporating this small variation of the mean value with temperature does not practically affect the results discussed below. The mean of the bond angle distribution does not change with temperature, as expected due to the equality of the $\alpha$ and $\zeta$ angles in the unstrained graphene and their constrained sum.

	%============================
	
	%============================

	A linear fitting describes the dependence of the variance $\Sigma^2_M$ of bond length distributions on temperature $T$
	\begin{equation}
		\label{eqn-CM}
		\Sigma^2_M(T) = C_M\, T,
		%1.6575081445565188
	\end{equation} 
	with $C_M = 1.66\times10^{-6}$~\AA$^2$/K [solid line in Figure~\ref{fig-zero_stress_distros_variances}(c)].  Similarly, the numerically found variances $\Sigma^2_B$ for the bond angle distributions are well described by 
	\begin{equation}
		\label{eqn-CB}
		\Sigma^2_B(T) = C_B\, T,
		%1.029378735528506,
	\end{equation} 
	with $ C_B = 1.02\times 10^{-2}$~deg$^2$/K [solid line in Figure~\ref{fig-zero_stress_distros_variances}(d)].
	
	The proportionality of these variances with temperature can be derived through the Boltzmann distribution when a quadratic approximation of the corresponding potential energy is considered.
	In particular, by the second derivative of the Morse potential of Equation~\eqref{eqn-morse}, 
	\begin{equation}
		V_M''(r) = -2Da^2e^{-2a(r-r_0)} \left( e^{a(r-r_0)}-2 \right),
		\label{eqn-V''}
	\end{equation} 
	the second order approximation of the bond stretching energy term about the equilibrium $r=r_0$ reads $V_M^{lin} = a^2D(r-r_0)^2$.
	Using this approximation, the corresponding Boltzmann distribution $\exp\left(\frac{-a^2D(r-r_0)^2}{k_BT}\right)$  results in a normal bond length distribution of the form $\exp\left(\frac{-(r-r_0)^2}{2\Sigma_M^2}\right)$
	%\begin{equation}
	%    \exp\left(\frac{-\varepsilon_i}{k_BT}\right) = \exp\left(\frac{-a^2D(r-r_0)^2}{k_BT}\right) = \exp\left(\frac{-(r-r_0)^2}{2\Sigma_M^2}\right),
	%\end{equation} 
	centered about the mean $r_0$ with a variance
	\begin{equation}
		\Sigma^2_M(T) = \frac{k_B}{2a^2D}\, T,
		\label{eq:S_M}
	\end{equation} 
	which gives $\Sigma^2_M(T)=1.97\times 10^{-6}$ (\AA$^2$/K) $\cdot \; T$ when the parameters of the Morse potential are substituted.

	Following a similar approach to estimate the variance of the bond angle distributions for different temperatures, we consider the second-order approximation of the potential $V_B$, Equation~\eqref{eqn-bending}, about $\phi=\phi_0=120^\circ$, given by $V_B^{lin}= \frac{d}{2}\frac{\pi^2}{180^2}(\phi-\phi_0)^2$ (when angles are measured in degrees). Note that due to the constrains in the sums of the bond angles around a particular atom and also in the sums within hexagonal rings, just one angle can not be varied alone. When an angle slightly varies from the equilibrium value, at least three other angles should be also changed. Thus, multiplying by a factor 4 the linearized angle bending energy in the Boltzmann distribution, we eventually find the variance (in squared degrees), about the mean $\phi_0$, of the bond angle distribution 
	\begin{equation}
		\Sigma^2_B(T) = \frac{k_B180^2}{4\pi^2d}\, T,
		\label{eq:S_B}
	\end{equation} 
	which results in $\Sigma^2_B(T)=1.01\times 10^{-2}$ (deg$^2$/K) $\cdot \, T$, using the value of $d$.
	
	Dotted lines in Figures~\ref{fig-zero_stress_distros_variances}(c) and (d) correspond to the analytical expressions of Equations~\eqref{eq:S_M} and \eqref{eq:S_B}, respectively.
	We can see from Figure~\ref{fig-zero_stress_distros_variances}(c) that the analytically obtained slope of Equation~\eqref{eq:S_M} is somehow larger than the corresponding numerical value of Equation~\eqref{eqn-CM} (the relative difference is less than 20\%). Concerning the variance of the bond angle distributions, Figure~\ref{fig-zero_stress_distros_variances}(d) shows an excellent agreement between the analytically and numerically obtained slopes, exhibiting a relative difference of less than 1\%.
	One reason for the quantitative disparity between the analytical prediction and numerical determination of the slope in the linear temperature dependence of the variance of the bond length distributions, but not for the angle distributions, may be due to the fact that the second order approximation of the angle bending potential $V_B(\phi)$ of Equation~\eqref{eqn-bending} is valid for a wide range of angles (see Figure~2 in Reference~\cite{ref-Kalosakas-JAP-2013}). On the other hand, due to the highly anharmonic nature of the Morse potential $V_M(r)$, Equation~\eqref{eqn-morse}, in the same energy scales (see Figure~1 in Reference~\cite{ref-Kalosakas-JAP-2013}), the second order approximation about $r_0$ is only valid very close to $r_0$.
	%Thus for small temperatures, for which the achieved bond lengths are all nearby $r_0$, the analytical and numerical predictions in Figure~\ref{fig-zero_stress_distros_variances}(c) are quite close. Nevertheless, as the temperature $T$ increases, much larger bond lengths come into play and hence the analytical prediction of the variances significantly deviates from the numerical results.
	
	Combining now the numerically determined variances for different temperatures and the known bond length and angle mean values as a function of the applied stress/strain, analytical approximate expressions for the bond length and angle distributions can be derived.
	Regarding the bond length distributions, an additional issue should be taken into account when the numerically determined variances from Equation~\eqref{eqn-CM} will be used. In particular, the relation between the variance and the temperature should be scaled according to the behavior of $V''_M$, Equation~\eqref{eqn-V''}, at the mean of the corresponding peak of the distribution, since bond lengths even further away from $r_0$ are encountered once stress is applied to the system and the second derivative of the Morse potential varies significantly with $r$. Given that analytically the variance equates to $k_BT/V_M''(r_0)$ close to $r=r_0$, we multiply the numerically determined variance from Equation~\eqref{eqn-CM} by the scaling function
	\begin{equation}
		\label{eqn-F}
		F(r) = \frac{V_M''(r_0)}{V_M''(r)},
	\end{equation}
	where $r$ is the known mean of the peak of interest in the distribution, provided by either Equations~\eqref{eqn-ac-lengths-A} - \eqref{eqn-ac-lengths-Z} or Equations~\eqref{eqn-zz-lengths-A} - \eqref{eqn-zz-lengths-Z} depending on the loading direction.
	
	As a result, the bond length distribution for a given applied stress/strain and temperature $T$ can be approximated by the relation
	%\begin{adjustwidth} {-\extralength}{0cm}
	\begin{equation} 
		P_M=\frac{1}{3\sqrt{2\pi \, C_M F\left(A\right)\, T}}\exp\left(-\frac{(r-A)^2}{2\, C_M F\left(A\right)\, T}\right) + \frac{2}{3\sqrt{2\pi\, C_M F\left(Z\right)\, T}}\exp\left(-\frac{(r-Z)^2}{2\, C_M F\left(Z\right)\, T}\right),
		\label{eqn-M_fitting}
	\end{equation}
	%\end{adjustwidth} 
	where $A$ and $Z$ are functions of the applied stress/strain, determined in Equations~\eqref{eqn-ac-lengths-A} - \eqref{eqn-ac-lengths-Z} or Equations~\eqref{eqn-zz-lengths-A} - \eqref{eqn-zz-lengths-Z}, for stress along the armchair or zigzag direction respectively, $C_M$ is given in Equation~\eqref{eqn-CM}, and $F(r)$ is obtained by Equation~\eqref{eqn-F}. The factor of 2 in the second term is because there are double $Z$ bonds than $A$ bonds. The division by 3 is for normalizing the distribution.
	Note that the quantities $A$ and $Z$ are provided by the corresponding zero temperature relations in Equations~\eqref{eqn-ac-lengths-A}, \eqref{eqn-ac-lengths-Z}, \eqref{eqn-zz-lengths-A}, and \eqref{eqn-zz-lengths-Z}  as a function of strain $\epsilon$. If they are needed as a function of stress $\sigma$, the stress-strain relation of Equation~\eqref{eqn-stress-strain} should be used to change the variable of the applied load.

	For the bond angle distributions, the subtlety mentioned above concerning the scaling function is not needed since the second derivative of the angle bending potential, Equation~\eqref{eqn-bending}, is everywhere the same regardless of the angle value at the peak of the distribution. Therefore, the angle bending distributions can be approximated by the expression
	\begin{equation}
		P_B = \frac{1}{3\sqrt{2\pi\,  C_B \, T}}\left[\exp\left(-\frac{(\phi-\alpha)^2}{2\, C_B \, T}\right) + 2\,\exp\left(-\frac{(\phi-\zeta)^2}{2\, C_B \, T}\right)\right],
		\label{eqn-B_fitting}
	\end{equation}
	where $\alpha$ and $\zeta$ are determined by the applied stress/strain from Equations
	\eqref{eqn-ac-angles-alpha} - \eqref{eqn-ac-angles-zeta} or \eqref{eqn-zz-angles-alpha} - \eqref{eqn-zz-angles-zeta}, depending on the loading direction, and $C_B$ is provided by Equation~\eqref{eqn-CB}. {The factor 2 in the second term is due to there being twice as many $\zeta$ angles as $\alpha$ angles and the division by 3 normalizes the distribution.} If the loading is given through the value of stress, Equation~\eqref{eqn-stress-strain} can be also used.
	
\begin{figure}[H]
		\centering
%		\begin{adjustwidth}{-\extralength}{0cm}
			\centering{\includegraphics[width=16cm]{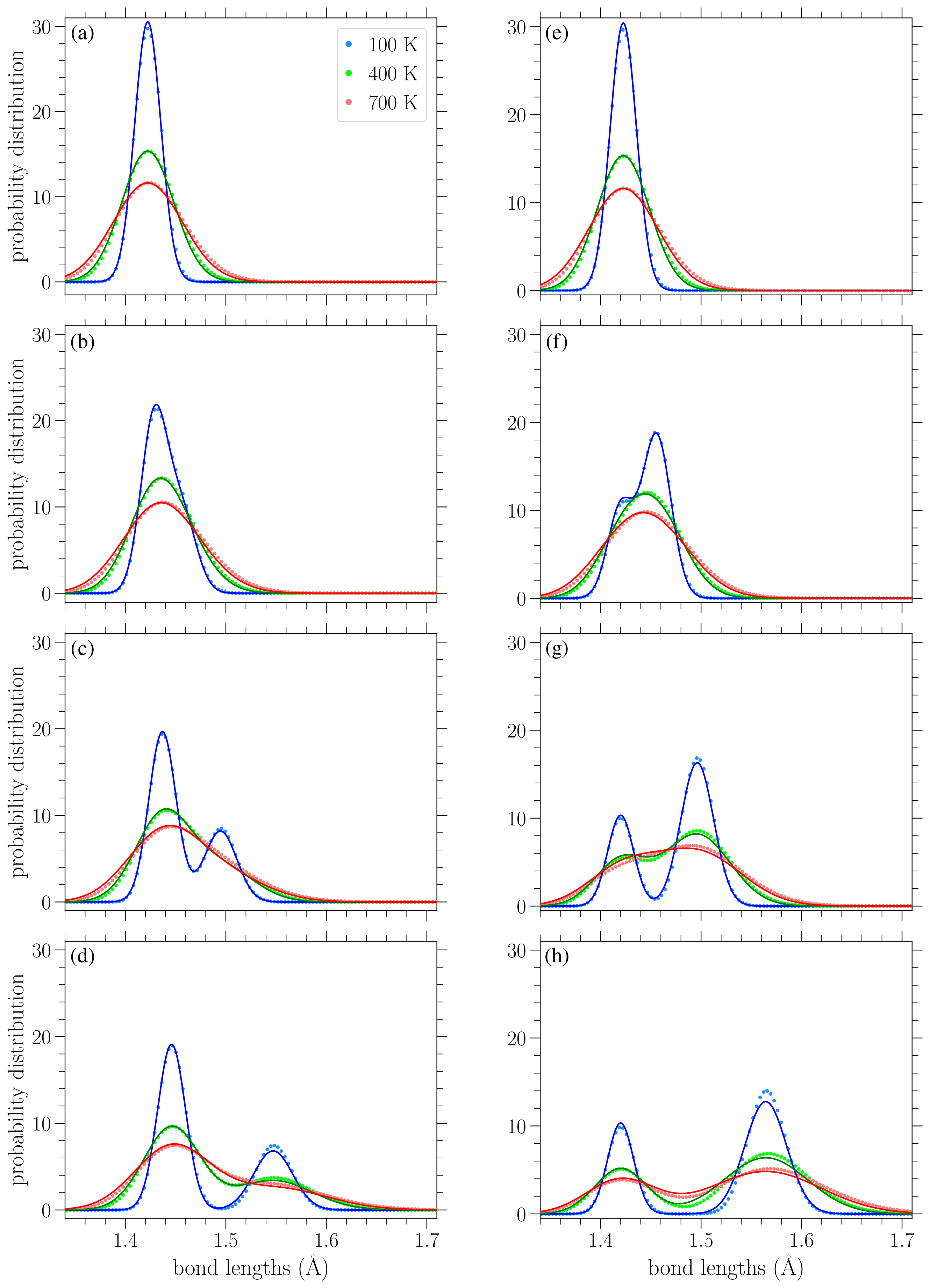}}
%		\end{adjustwidth}
		
		\caption{Bond length distributions at various temperatures $T$ (as shown in the legend) and applied stress along the armchair direction (left column) or the zigzag direction (right column), for increasing loads from top to bottom. In particular, the stresses $\sigma$ in the armchair direction are \textbf{(a)} 0.0813~eV/\AA$^2$, \textbf{(b)} 0.569~eV/\AA$^2$, \textbf{(c)} 1.06~eV/\AA$^2$, and \textbf{(d)} 1.55~eV/\AA$^2$. The stresses along the zigzag direction are \textbf{(e)} 0.0939~eV/\AA$^2$, \textbf{(f)} 0.751~eV/\AA$^2$, \textbf{(g)} 1.41~eV/\AA$^2$, and \textbf{(h)} 2.16~eV/\AA$^2$. The analytical expressions of Equation~\eqref{eqn-M_fitting} are shown by solid curves and the corresponding numerical data by circles.}
		\label{fig-lengths_fitted}
		%============================
		%============================
	\end{figure}

\begin{figure}[H]
		\centering
%		\begin{adjustwidth}{-\extralength}{0cm}
			\centering{\includegraphics[width=16cm]{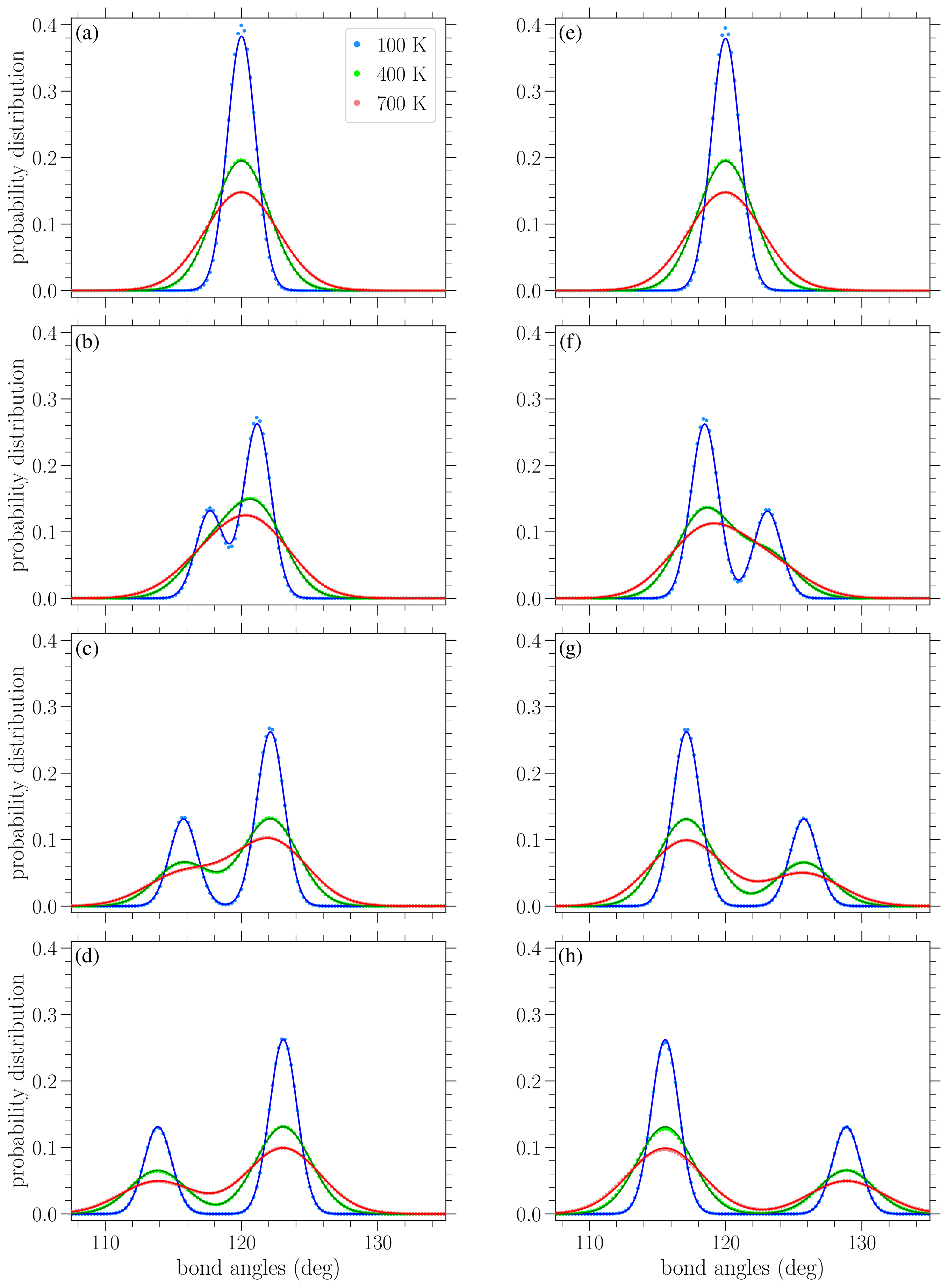}}
%		\end{adjustwidth}
		
		\caption{Bond angle distributions for different temperatures $T$ (as shown in the legend) and stress along the armchair direction (left column) or the zigzag direction (right column), for increasing values of stress from top to bottom. The stress in the armchair direction is \textbf{(a)} 0.0813~eV/\AA$^2$, \textbf{(b)} 0.569~eV/\AA$^2$, \textbf{(c)} 1.06~eV/\AA$^2$, and \textbf{(d)} 1.55~eV/\AA$^2$, while along the zigzag direction is \textbf{(e)} 0.0939~eV/\AA$^2$, \textbf{(f)} 0.751~eV/\AA$^2$, \textbf{(g)} 1.41~eV/\AA$^2$, and \textbf{(h)} 2.16~eV/\AA$^2$. The analytical expressions given by Equation~\eqref{eqn-B_fitting} are indicated by the solid curves and the corresponding numerical data by circles.}
		\label{fig-angles_fitted}
	\end{figure}

	Circles in Figure~\ref{fig-lengths_fitted} present the numerically computed bond length distributions at various applied stresses and temperatures, while the solid lines correspond to the curve $P_M$ from Equation~\eqref{eqn-M_fitting}. Figure~\ref{fig-angles_fitted} contains similar results, but for the bond angle distributions. 
	These plots show that the analytical expressions presented above provide overall a reasonable description of the bond length and angle distributions in strained graphene, at various temperatures at least up to the values considered here.
	In Figure~\ref{fig-lengths_fitted}, at the larger values of applied stress and lower temperatures the analytical distribution $P_M$ of Equation~\eqref{eqn-M_fitting} underestimates the longer-bond (second) peak of the numerically obtained distributions  [Figures~\ref{fig-lengths_fitted}(d), (g) and (h)]. {In Figure~\ref{fig-lengths_fitted}(h) we observe the greatest deviation of the analytical expression from the numerical data at the right-hand peak of the lowest temperature at $T=100$~K; in this case the difference is 8.7\%.}
	From the plots of Figure~\ref{fig-angles_fitted} we see that the analytical expression $P_B$, Equation~\eqref{eqn-B_fitting},  describes the data quite well, apart from small discrepancies at the heights of the taller peak at the lower temperatures depicted and for the smaller values of stress. {In Figure~\ref{fig-angles_fitted}(a) we observe the biggest deviation for the $T=100$~K case, where the value of the analytical expression is 4.1\% below that of the numerical data.} In any case, both expressions of Equations~\eqref{eqn-M_fitting} and \eqref{eqn-B_fitting} provide a useful analytical description of the underlying structural properties of the strained graphene at finite temperatures.

	%============================
	% \todo[inline]{we display f = 0.1, 0.7, 1.3, 1.9 for ac dir; f = 0.1, 0.8, 1.5, 2.3 for zz dir}
	%============================

	%%%%%%%%%%%%%%%%%%%%%%%%%%%%%%%%%%%%%%%%%%
	\section{Conclusions}
	\label{section-conclusion}
	We investigated the planar dynamics of a uniaxially loaded graphene sheet using Hamiltonian formalism and an efficient symplectic integration technique allowing the creation of accurate numerical data for very long simulation times. Our MD simulations examined the effects of thermal fluctuations in the mechanical response of graphene. In particular, we derived stress-strain responses for two different directions of applied stress, along either the armchair or the zigzag direction, at various temperatures. 
	
	A small, almost linear decrease of the {Young} modulus of graphene is obtained as the temperature of the sheet increases. Such a variation of {Young} modulus with temperature is in line with previous investigations. Furthermore, an intriguing temperature dependence of the third-order elastic modulus {has been presented for the first time}, which is found to decrease (slightly increase) its absolute value with increasing temperature, for stresses along the armchair (zigzag) direction. Finally, we found that the tensile strength and failure strain decrease approximately linearly with temperature and computed the slope of this variation. {A quantitative comparison with existing results regarding these variations is presented.}
	
	{It is worth mentioning that even though our model is restricted to planar deformations, results obtained for the intrinsic strength and fracture strain at room temperature are in agreement with MD simulations \cite{ref-Zhao-NanoLet-2009,ref-Varillas-MechSci-2024, ref-Lee-NanoResLets-2015} allowing out-of-plane displacements of carbon atoms.} {Moreover, the values of Young modulus and fracture strength at 300~K are in accordance with experimental estimates presented in References \cite{ref-Lee-Science-2008, ref-Cao-NatComm-2020, ref-Varillas-MechSci-2024} and Reference~\cite{ref-Lee-Science-2008}, respectively.}
	
	The dependence of the distributions of bond lengths and bond angles within the graphene sheet, on both the applied stress and temperature is also discussed. Approximate analytical expressions for these distributions are provided. In particular, we found that the distributions can be described by the sum of two Gaussian peaks, where the center of each peak is obtained from the values of bond lengths or bond angles, respectively, in the strained graphene subjected to the particular amount of stress at zero temperature. The variance of each peak as a function of temperature can be derived by the corresponding data at zero applied stress, while for the bond length distributions a scaling factor is additionally incorporated to account for the anharmonicity of the Morse potential. 
	Thus, {for the first time, a detailed description of the effects of both stress and temperature on the structural properties of graphene is reported.}

		{\textbf{Acknowledgements:} S.~E. acknowledges support from the National Research Foundation of South Africa, the University of Cape Town (UCT), and the Science Faculty of UCT. All authors thank the Erasmus+/International Credit Mobility KA171 program for support. We thank the Centre for High Performance Computing (CHPC) of South Africa for providing the computational resources for this project. S.~E. thanks M.~Hillebrand, B.~Many Manda, and A.~Ngapasare for their insight regarding the numerical simulations, and D.~Kruyt for their technical assistance with code optimization and systems administration.}
		
		{\textbf{Author contributions:} Conceptualization, Ch.~S. and G.~K.; methodology, Ch.~S. and G.~K.; software, S.~E.; validation, S.~E., Ch.~S. and G.~K.; formal analysis, S.~E., Ch.~S. and G.~K.; investigation, S.~E., Ch.~S. and G.~K.; writing---review and editing, S.~E., Ch.~S. and G.~K.; visualization, S.~E.; supervision, Ch.~S. and G.~K. All authors have read and agreed to the published version of the manuscript.}

		\bibliographystyle{unsrt}
		\bibliography{Paper_1_bib.bib}  
	\end{document}